\documentclass{sistedes}
\usepackage{graphicx}
\usepackage{subcaption}
\usepackage{color}
\usepackage{pgfplots}
\usepgfplotslibrary{groupplots}

\usepackage[a4paper,left=20mm,right=20mm,top=40mm,bottom=20mm]{geometry}
\usepackage{adjustbox}
\usepackage{amsmath,amsfonts,amssymb}
\usepackage{algorithmic}
\usepackage{array}
\usepackage{textcomp}
\usepackage{stfloats}
\usepackage{url}
\usepackage{verbatim}
\usepackage{cite}
\usepackage{balance}
\usepackage[titletoc]{appendix}
\usepackage{bbding}
\usepackage{pifont}

\DeclareGraphicsExtensions{.pdf,.jpeg,.png}
\usepackage{multirow}
\usepackage{kotex}
\usepackage{makecell}

\usepackage[noabbrev,nameinlink]{cleveref}
\usepackage{hyperref}
\hypersetup{
    colorlinks = false, 
    linkbordercolor = {1 1 1}, 
    citebordercolor = {1 1 1}, 
    urlbordercolor = {1 1 1}, 
}

\begin{document}
\title{Expanding the Attack Scenarios of SAE J1939: A Comprehensive Analysis of Established and Novel Vulnerabilities in Transport Protocol}

\author{Hwejae Lee\inst{1, 2} \and
Hyosun Lee\inst{1,3} \and
Saehee Jun\inst{1,4} \and
Huy Kang Kim\inst{1,5}}
\authorrunning{ }

\institute{
Korea University, Seoul, Korea\\
\and
\email{hwejae94@korea.ac.kr}\\
\and
\email{hslee@korea.ac.kr}\\
\and
\email{junsaehee@korea.ac.kr}\\
\and
\email{cenda@korea.ac.kr}
}

\maketitle              
\begin{abstract}
Following the enactment of the UN Regulation, substantial efforts have been directed toward implementing intrusion detection and prevention systems (IDPSs) and vulnerability analysis in Controller Area Network (CAN). However, Society of Automotive Engineers (SAE) J1939 protocol, despite its extensive application in camping cars and commercial vehicles, has seen limited vulnerability identification, which raises significant safety concerns in the event of security breaches. 
In this research, we explore and demonstrate attack techniques specific to SAE J1939 communication protocol.
We introduce 14 attack scenarios, enhancing the discourse with seven scenarios recognized in the previous research and unveiling seven novel scenarios through our elaborate study.
To verify the feasibility of these scenarios, we leverage a sophisticated testbed that facilitates real-time communication and the simulation of attacks.
Our testing confirms the successful execution of 11 scenarios, underscoring their imminent threat to commercial vehicle operations.
Some attacks will be difficult to detect because they only inject a single message.
These results highlight unique vulnerabilities within SAE J1939 protocol, indicating the automotive cybersecurity community needs to address the identified risks.

\end{abstract}
\section{Introduction} \label{sec:introduction}

Recently, security issues in the automotive industry have emerged as a significant challenge that can no longer be overlooked\cite{risingSecurityWeaknesses}.
Especially for commercial vehicles, these security issues are even more serious.
Commercial vehicles require special attention in terms of security because the damage from accidents is greater compared to passenger cars\cite{nmfta1}.
This is particularly the case for large commercial vehicles such as trucks and buses due to their size and the importance of the cargo they carry from a security standpoint\cite{nmfta2}.

Interest in commercial vehicle security is showing an upward trend, as is the active organization of hacking challenges targeting commercial vehicles such as the CyberTruck Challenge\cite{cybertruckchallenge}. Also, interest in commercial vehicles has surged as autonomous and platooning technologies have advanced.
These technologies are more crucially utilized in commercial vehicles than in general vehicles, and the security requirements associated with them require special consideration\cite{hackingThreatsinCA}.
SAE J1939 protocol is widely adopted for commercial vehicles, while CAN is used for passenger vehicles. In addition, various data, including the parameter group number (PGN) and data fields, are also standardized and open to the public. 

However, current security research is mainly focused on passenger cars, resulting in a relative lack of specialized research for commercial vehicles using SAE J1939.
This implies that there may be security vulnerabilities in commercial vehicles using SAE J1939, and such vulnerabilities can impact various industries where commercial vehicles play a crucial role.
Therefore, commercial vehicles, having different characteristics and requirements from passenger cars, need a specialized security approach that takes these factors into account.
This research focuses on SAE J1939 protocol from this point of view. 
The characteristics of this protocol, widely used in commercial vehicles, expand from those of the CAN protocol, making it essential to consider security threats and attack scenarios specific to commercial vehicles.
This paper aims to provide an in-depth understanding and a new approach in the field of commercial vehicle security by exploring the concepts and feasibility of attack scenarios specialized for SAE J1939 protocol in commercial vehicles.
It emphasizes that responding to security threats in commercial vehicles goes beyond a mere technical challenge, as it is directly linked to national safety.

The remainder of this paper is organized as follows.
In Section \ref{sec:Background}, we present an overview of SAE J1939 protocol, which serves as a fundamental standard for inter-component communication in commercial vehicles. The section elaborates on the protocol's architecture, various message types, and the specialized procedures for multi-packet data exchange, along with the functionalities offered by the protocol for commercial vehicle communication systems.
In Section \ref{sec:related work}, we present an overview of attack scenarios examined in the existing CAN framework alongside those explored within SAE J1939 standard.
Section \ref{sec:experiment testing setup} details our testbed environment, while Section \ref{sec:experiment} delineates a total of 14 attack scenarios, encompassing both those identified in prior research and new scenarios proposed within this testbed. Finally, Section \ref{sec:conclusion} concludes with a summary of the findings and outlines directions for future work.
\section{Background} \label{sec:Background}

SAE J1939 stipulates the recommended practices for trucks and buses regarding communication between internal components in commercial vehicles. The standard originated in the automobile and heavy truck industries in the United States, has recently spread, and is now widely adopted worldwide.
These protocols are not only used for internal communication in cargo trucks or trailers but are also widely applied in commercial vehicles in various industries (e.g., construction machinery, agricultural machinery, ships, military vehicles), implying that vulnerabilities in SAE J1939 standard may be common to a wide range of commercial vehicle models. 

\begin{figure}[!t]
    \centering
    \begin{tabular} {c}\\
    \includegraphics[width=0.99\textwidth]{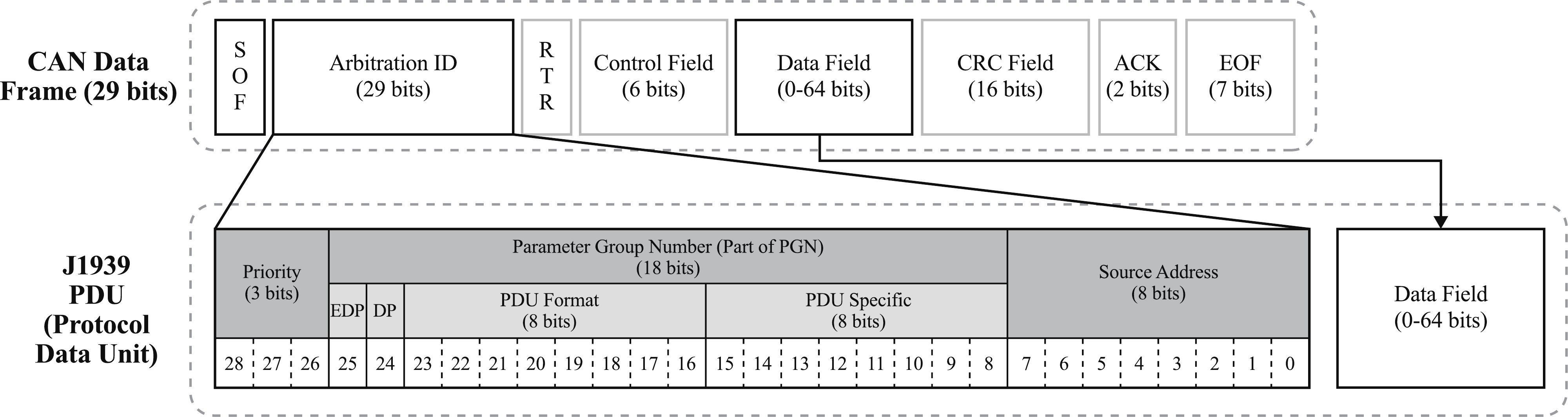} \\
    \end{tabular}
    \caption{Data unit format of SAE J1939 protocol}
    \label{fig:J1939_character_0}
\end{figure}

SAE J1939 standard supports a higher-level protocol that enables manufacturers of large industrial electronic control units (ECUs) to communicate over large, often complex networks. SAE J1939 application protocol uses a 29-bit CAN extended frame identifier. The arbitration identifier (AID) is divided into several parts, including the PGN, which identifies the frame and defines which signals are included. The first three bits of the ID are the priority bits. What makes up the next 18 bits is the PGN. Each frame can be transmitted to a global address or a specific address, and the PGN includes the following related information: extended data page (EDP), data page (DP), protocol data unit (PDU) format (PF), and PDU specific (PS). The last eight bits of the frame contain the source address (SA). \Cref{fig:J1939_character_0} shows the data unit format of SAE J1939 protocol.
\begin{figure}[!t]
  \centering

  \begin{subfigure}{.33\textwidth}
    \centering
    \includegraphics[width=0.65\linewidth]{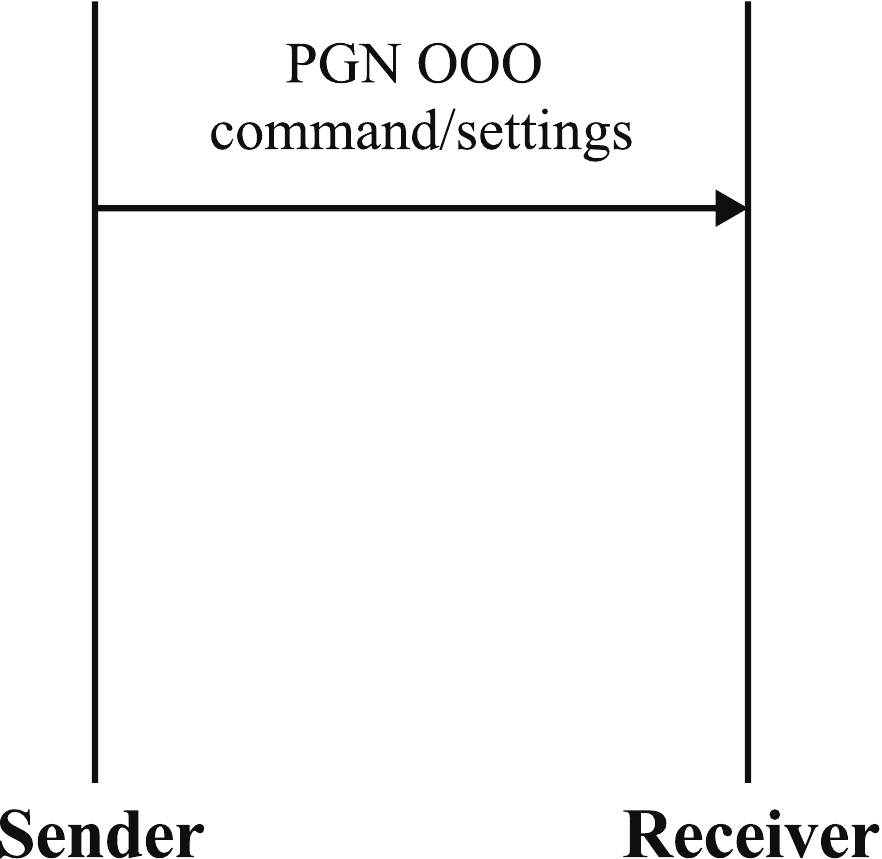}
    \caption{Command message\\ {$\ $}}
    \label{fig:J1939_character_1}
  \end{subfigure}%
  \begin{subfigure}{.33\textwidth}
    \centering
    \includegraphics[width=0.65\linewidth]{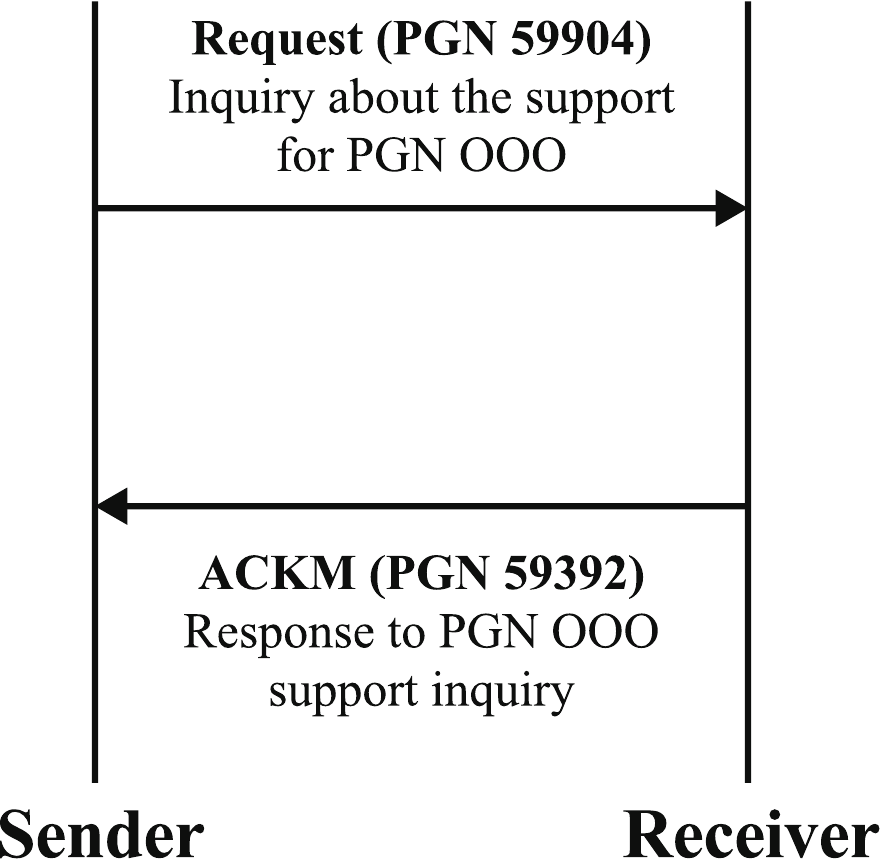}
    \captionsetup{justification=centering}
    \caption{Request and\\Acknowledgement message}
    \label{fig:J1939_character_2}
  \end{subfigure}%
  \begin{subfigure}{.33\textwidth}
    \centering
    \includegraphics[width=0.65\linewidth]{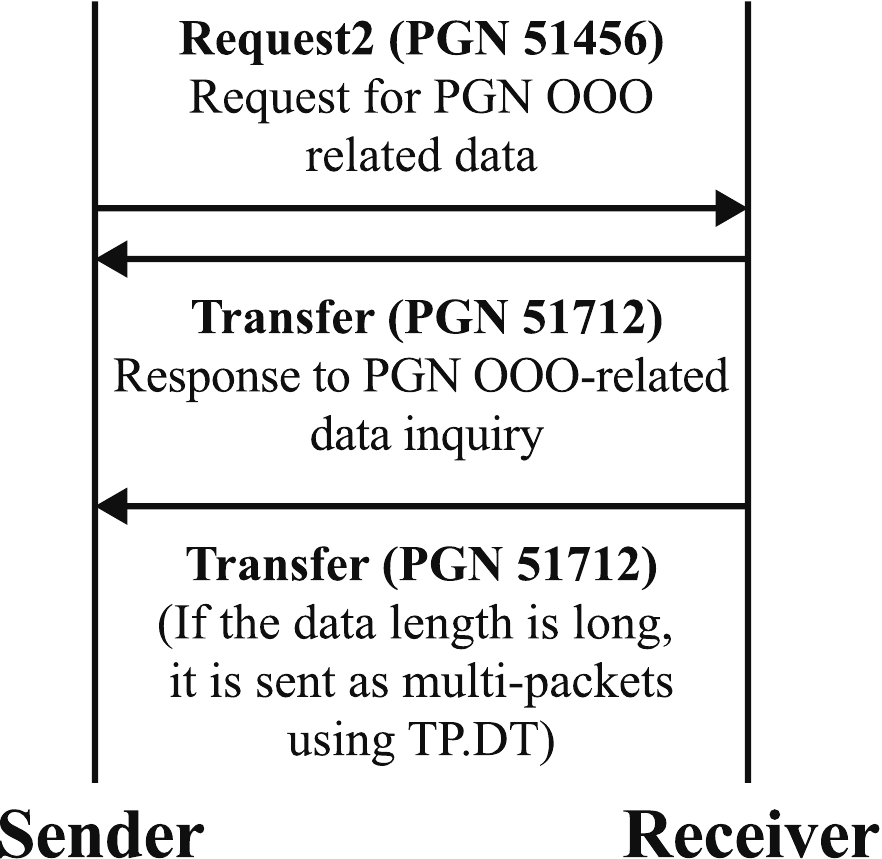}
    \caption{Request2 and Transfer message\\ {$\ $}}
    \label{fig:J1939_character_3}
  \end{subfigure}%
  
  \begin{subfigure}{.55\textwidth}
    \centering
    \begin{tabular} {c}\\
    \includegraphics[width=.95\textwidth]{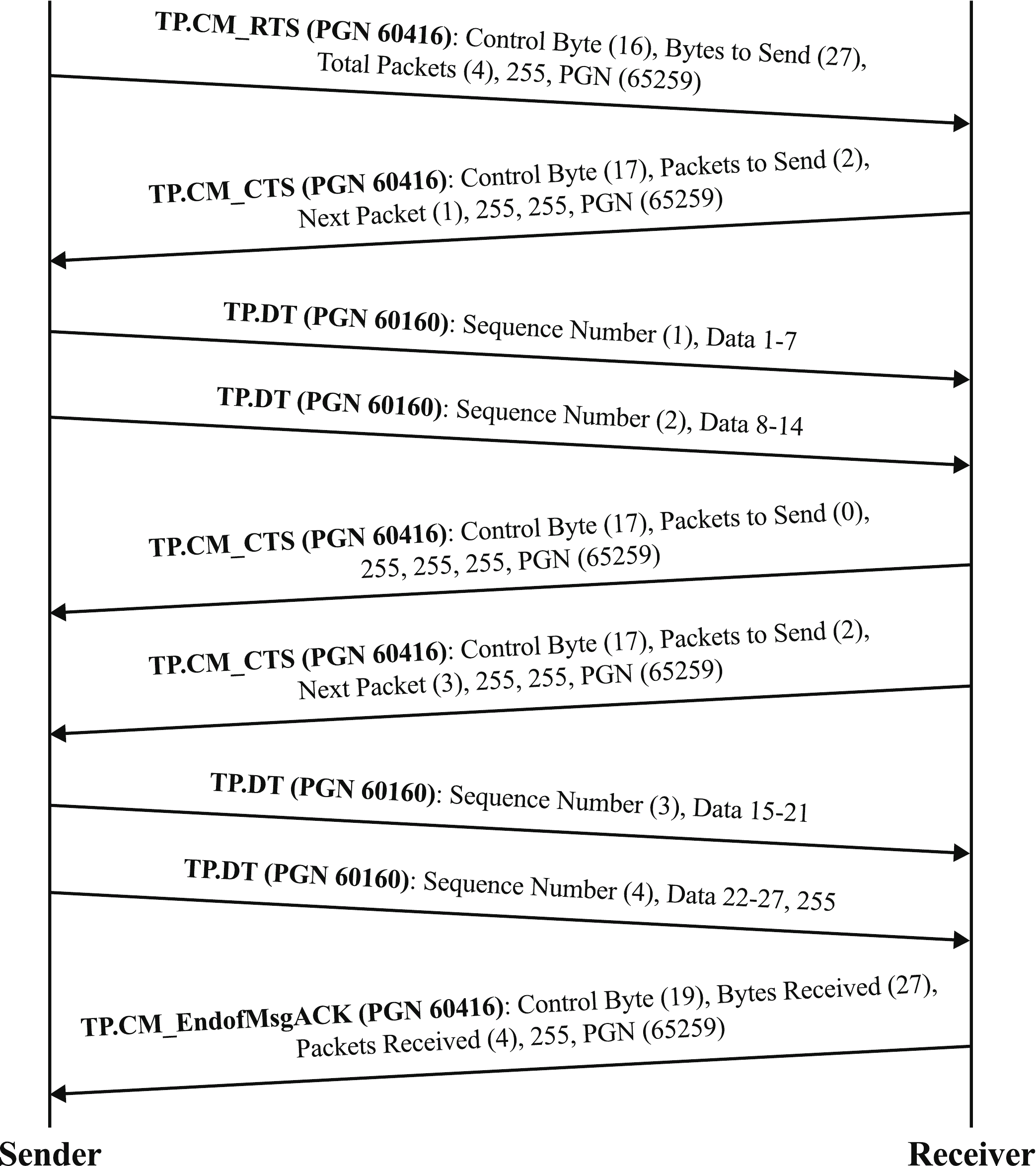} \\
    \end{tabular}
    \caption{Transport protocol transfer sequences}
    \label{fig:J1939_character_4}
  \end{subfigure}
  \caption{SAE J1939 message transmission sequences}
  \label{fig:J1939_character}
  
\end{figure}
PDU can be divided into PDU1 and PDU2 depending on the PF value. PDU1 is a message that follows the peer-to-peer method with a PF value from 0 to 239. In PDU1, PS refers to the destination address (DA). PDU2 corresponds to a message broadcast to the specified group with a PF value from 240 to 255. In this case, PS means group extension (GE).

The meaning and role of a SAE J1939 message can be interpreted through PGN. \Cref{fig:J1939_character} shows a typical communication method in SAE J1939. A Command (various PGNs) message is a message in which the sender commands the receiver to perform a specific operation such as torque and speed control. This receiver is defined in the parameter group. A Request (PGN 59904) and an Acknowledgement (PGN 59392) messages form a pair. The sender can query the receiver as to whether to use a specific PGN using the Request message. The receiver sends the Acknowledgement message in response. Lastly, there are Request2 (PGN 51456) and Transfer (PGN 51712) messages. When the receiver receives a request for data related to a specific PGN through a Request2 message, the response is sent as a Transfer message.
Multi-packet data transmission is possible in SAE J1939 standard communication. A connection setup process must precede it. This process is based on a transport protocol connection management (TP.CM, PGN 60416) message. There are five types of TP.CM messages: request to send (TP.CM\_RTS), clear to send (TP.CM\_CTS), end of message acknowledgment (TP.CM\_EndOfMsgACK),  connection abort (TP.Conn\_Abort), and broadcast announce message (TP.CM\_BAM). The type can be distinguished by the control byte (16, 17, 19, 255, 32), which is the first byte of the message. The first four types are messages involved in establishing a connection when transmitting multi-packet data.
During the connection initiation process, the sender transmits a TP.CM\_RTS message to the receiver. In TP.CM\_RTS, the SA field corresponds to the sending node and the DA field corresponds to the receiving node. The message includes the total message size, the total number of packets, and the PGN of the packeted message. Additionally, the maximum number of packets of a single TP.CM\_CTS to be received from one receiving node can be limited. If the same TP.CM\_RTS message is received multiple times from the same SA, the most recent message is applied and the previous messages are ignored.
In the connection established process, in contrast, the receiver sends a TP.CM\_CTS message, a connection acceptance message, to the sender in response to the TP.CM\_RTS message. The TP.CM\_CTS message contains the number of packets that can be sent, the next packet number to be sent, and the PGN of the packeted message. At this time, the size of the message to be sent must not exceed the maximum size as stipulated by TP.CM\_RTS. If multiple TP.CM\_CTS messages are received while the connection established is completed, the connection is interrupted. If a timeout occurs, the responder can terminate the connection or send a TP.CM\_CTS message to maintain the connection.
Subsequently, the sender transmits packets according to the number of packets requested. A large message is divided into several individual transport protocol data transfer (TP.DT, PGN 60160) packets and transmitted. Global DA (255) is only allowed in TP.CM\_BAM messages cannot be used during general TP.CM\_RTS/CTS multi-packet data transmissions.
Finally, when the receiver receives all packets, it sends TP.CM\_EndOfMsgACK message to inform the sender that the entire message has been successfully received and the connection is terminated. The TP.CM\_EndofMsgACK message contains the total message size, the number of packets the receiver has received, and the PGN of the packeted message.
However, if the sender or receiver wants to close the connection for any reason for disconnection, they can stop the connection by sending a TP.Conn\_Abort message. The message includes the connection abort reason, the role of the sender, and the PGN of the packeted message.
A TP.CM\_BAM is used to notify all nodes in the network that a message will be broadcast. This message includes the total message size, the total number of packets, and the PGN of the packeted message. After the TP.CM\_BAM message is transmitted, the TP.DT messages are broadcast, setting with the Global DA (255).
\section{Related Work} \label{sec:related work}

\begin{table*}[t]
\setlength{\tabcolsep}{5pt}
\caption{Various attacks on CAN protocols and related research.}
\resizebox{\textwidth}{!}{
\begin{tabular}{l|l|l}
\hline
\textbf{Attack Type} & \textbf{Description}                                                  &  \textbf{Authors}\\ \hline
Fuzzing     & send random data to the CAN bus, causing system errors. & \cite{martinelli2017car}, \cite{larson2008approach}, \cite{verma2020addressing}, \cite{seo2018gids}, \cite{jeong2023XCANIDS} \\ 
DoS         & overwhelms the network with excessive messages, disrupting its normal functions. & \cite{martinelli2017car}, \cite{larson2008approach}, \cite{verma2020addressing}, \cite{seo2018gids}, \cite{studnia2013survey}, \cite{song2016intrusion} \\ 
Spoofing    & involving the transmission of false messages to manipulate the behavior of the vehicle. & \cite{martinelli2017car}, \cite{larson2008approach}, \cite{verma2020addressing}, \cite{seo2018gids}, \cite{jeong2023XCANIDS}, \cite{song2016intrusion}, \cite{iehira2018spoofing} \\ 
Replay      & deceives the network by resenting captured legitimate messages. & \cite{larson2008approach}, \cite{jeong2023XCANIDS}, \cite{song2016intrusion}\\ 
Suspension  & interferes with system functionality by blocking or delaying specific messages. & \cite{larson2008approach}, \cite{verma2020addressing}, \cite{jeong2023XCANIDS}, \cite{palanca2017stealth}\\ 
Masquerade  & alters the data part of messages to disguise malicious intent. & \cite{larson2008approach}, \cite{verma2020addressing}, \cite{jeong2023XCANIDS}, \cite{iehira2018spoofing} \\ \hline
\end{tabular} }
\label{tab:CANattackResaerchtable}
\end{table*}

\subsection{CAN protocol Attack Scenarios}
The CAN protocol is a core communication technology extensively used in vehicle networks in the automotive and other industrial sectors. 
This protocol enables efficient exchanges of data between various ECUs within vehicles, thereby enhancing vehicle performance and safety.
However, the CAN protocol inherently lacks security features, posing the risks of network intrusions and data manipulation by external attackers.
Recent researchers are focusing on exploring various attack scenarios that exploit these vulnerabilities and on developing security measures to counteract them.

The CAN protocol within vehicles is vulnerable to various attacks, primarily classified as fuzzing, denial-of-service (DoS), spoofing, replay, suspension, and masquerade attacks.
Fuzzing attacks involve sending random data to the network to induce system errors.
This type can be used to explore system vulnerabilities or as a stability test.
DoS attacks are carried out by sending excessive messages to the network, disrupting the normal functions of the vehicle.
Spoofing attacks manipulate vehicle behavior by sending false messages, which may appear normal but actually convey incorrect information.
Replay attacks use retransmitted captured messages to deceive the network, allowing the execution of unwanted commands.
Suspension attacks disrupt the normal operation of vehicle systems by blocking or delaying specific messages.
Finally, masquerade attacks are carried out by altering the data portion of messages to disguise malicious content by making it appear legitimate. 
These attacks can be difficult to detect and are accomplished through subtle changes in the messages.
Research conducted on each attack technique is briefly presented in Table \ref{tab:CANattackResaerchtable}.

\subsection{SAE J1939 Attack Scenarios}
SAE J1939 standard is commonly applied to commercial vehicles in various industries and supports a high-level protocol for large-scale network communication. This standard uses a 29-bit CAN extended frame identifier to enable multi-packet data communication and peer-to-peer communication. These unique characteristics of SAE J1939 can provide a new attack vector different from that in the CAN communication environment. Based on this, studies have recently attempted to stage vehicle network attacks targeting commercial vehicles and discuss possible countermeasures against new attacks and the impacts of such attacks.

First, attack demonstration studies in SAE J1939 environment were conducted, inspired by network attacks in the existing CAN communication environment.
Burakove \textit{et al.} \cite{burakova2016truck} demonstrated an attack targeting three attack vectors (instrument panel, powertrain, retarder) in the application of SAE J1939 protocol. Attacks that could pose a significant threat to safety when a truck is driven were demonstrated, such as those that control the instrument panel gauges, randomly manipulate the engine RPM (Revolutions Per Minute), and disable engine braking. This attack demonstration was conducted on tractors and school buses, showing that SAE J1939-based attacks targeting large vehicles can commonly affect such vehicles from multiple industries.
Mukherjee \textit{et al.} \cite{mukherjee2016practical} demonstrated three DoS attacks that disrupt the normal operation of the vehicle by injecting messages into the bus. The three attacks were classified according to the message type (Request, TP.CM) and exploit type (implementation issues, specification issues) used in the attack.
Jichici \textit{et al.} \cite{jichici2023control} conducted an experiment based on a simulation model connecting the Simulink environment and the CANoe environment to address adversary actions and countermeasures in the control systems of large vehicles.
Recent studies have proposed and demonstrated original attacks by taking advantage of the fact that the characteristics of SAE J1939 protocol differ from those of the existing general CAN communication environment.
Murvay and Groza \cite{murvay2018security} suggested that vulnerabilities in SAE J1939 could lead to critical attacks on vehicle safety (impersonation, DoS, distributed DoS) and demonstrated attacks that could occur in SAE J1939 environment using a CANoe-based simulation tool. Three of the attacks demonstrated in their study were inspired by the fact that SAE J1939 performs address claims. One attack relies on the fact that the standard supports the transmission of data packets longer than eight bytes, and when transmitting multi-packets, the establishment of the type of connection takes place between the sender and the receiver. It was shown that a TP.Conn\_Abort request can be made after the connection. The remaining attack steps target the working set supported by the standard, but these were not demonstrated due to the limitations of the CANoe simulation.
Chatterjee \textit{et al.} \cite{chatterjee2023exploiting} conducted five attack demonstrations and undertook hypothesis verification targeting actual large vehicles. The first two scenarios verify previously proposed attack hypotheses, and the latter three scenarios present possible attack vectors that can be exploited in SAE J1939 network. Each attack test was conducted on four testbeds and in an actual truck environment. The three newly proposed attack scenarios are limited in that they succeeded in only one of the four testbeds, including the old engine control module (ECM), and failed to demonstrate the attack in the remaining three testbeds and in actual vehicles.

\section{Experiment Setup} \label{sec:experiment testing setup}

To validate the proposed attack scenarios, we employed Au SAE J1939 simulators (Generation II), integral in product development and validation processes, encompassing SAE J1939 product line testing, receipt inspection, and commercial demonstrations.
These simulators are capable of generating SAE J1939 signals for up to three controller applications related to the engine, anti-lock brake system (ABS), and transmission.
Furthermore, the Au SAE J1939 Message Center System (MCS) was utilized to facilitate comprehensive data transmission and reception configurations.
The simulation involves 27 PGNs representing a total of 234 signals, all of which communicate via TP, including Request and Acknowledgment messages.

\begin{figure}[!t]
    \centering
    \begin{tabular} {c}\\
    \includegraphics[width=0.99\textwidth]{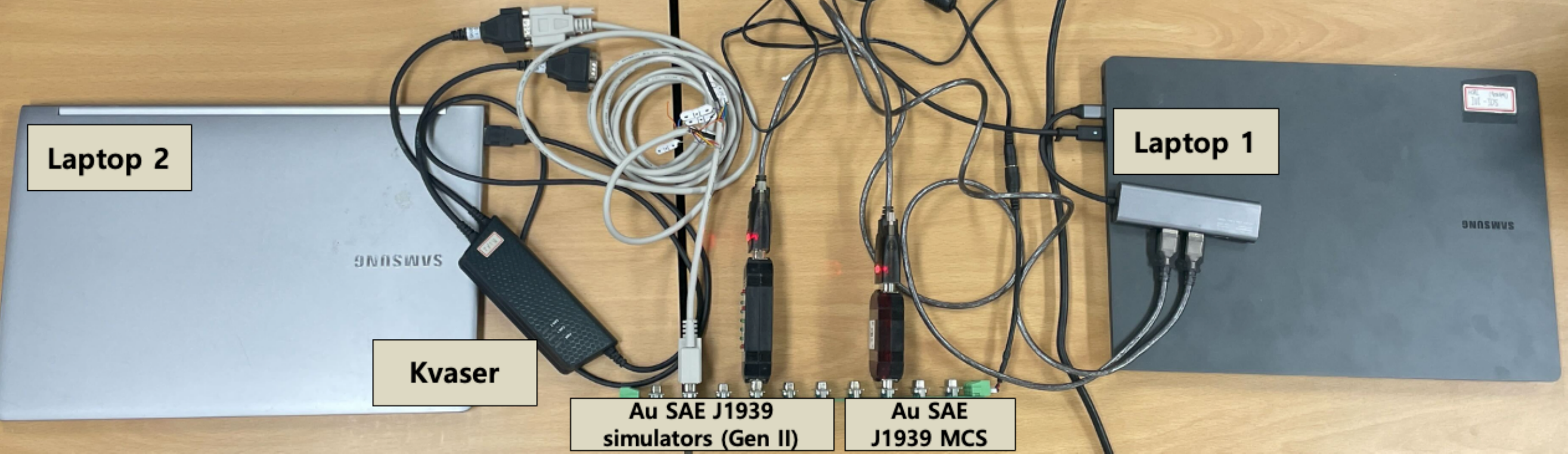} \\
    \end{tabular}
    \caption{J1939 attack simulation testbed setup. Laptop 1 monitors the CAN bus and Laptop 2 acts as the attacker within the SAE J1939 network, comprising Au SAE J1939 simulators (Generation II) and Au SAE J1939 MCS.}
    \label{fig:testbed}
\end{figure}

\begin{figure}[!t]
  \centering
  \begin{subfigure}{.49\textwidth}
    \centering
    \captionsetup{justification=centering}
    \includegraphics[width=0.9\linewidth]{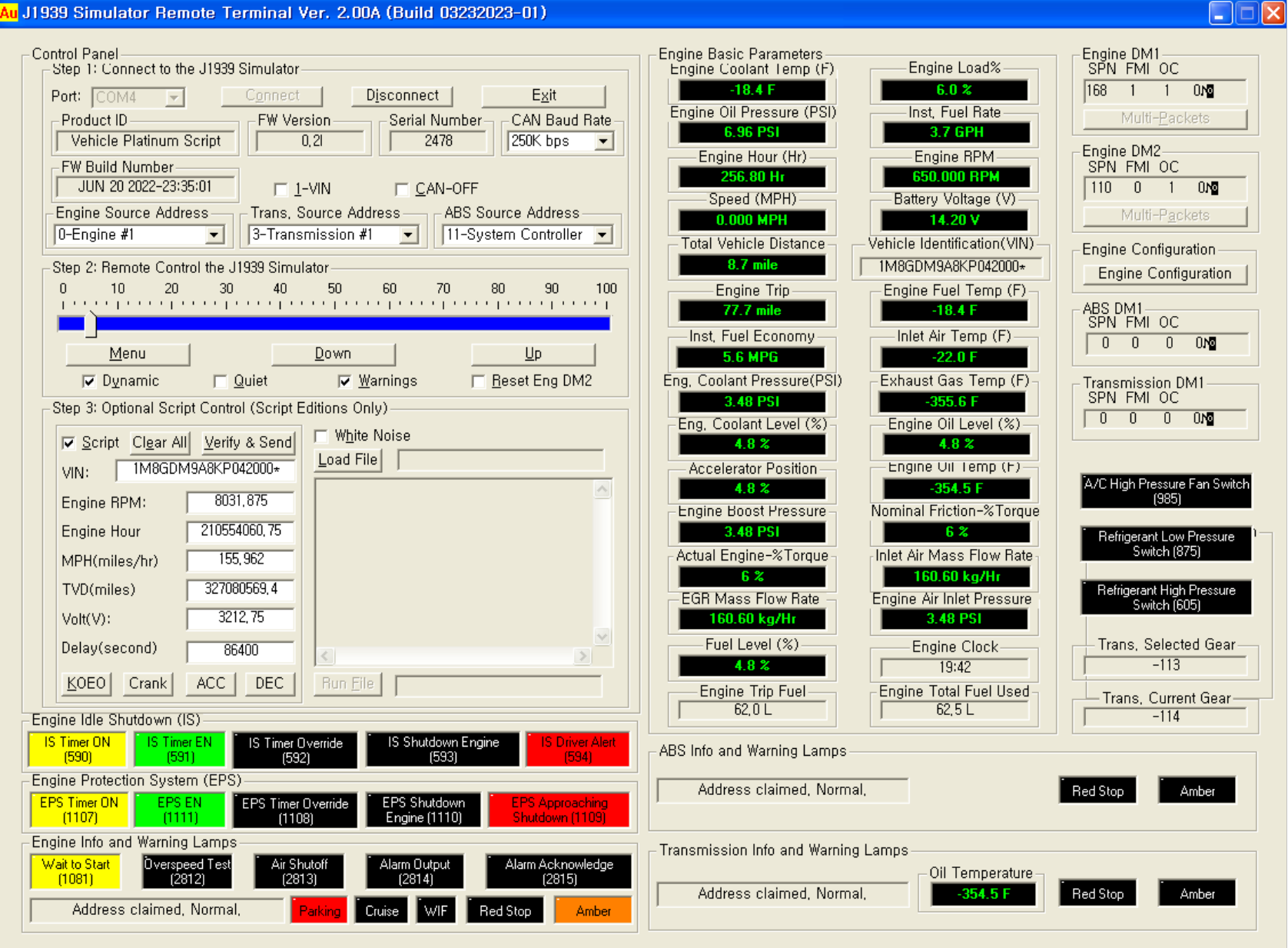}
    \caption{Au SAE J1939 simulator 2.00A remote terminal}
  \end{subfigure}%
  \begin{subfigure}{.49\textwidth}
    \centering
    \captionsetup{justification=centering}
    \includegraphics[width=0.9\linewidth]{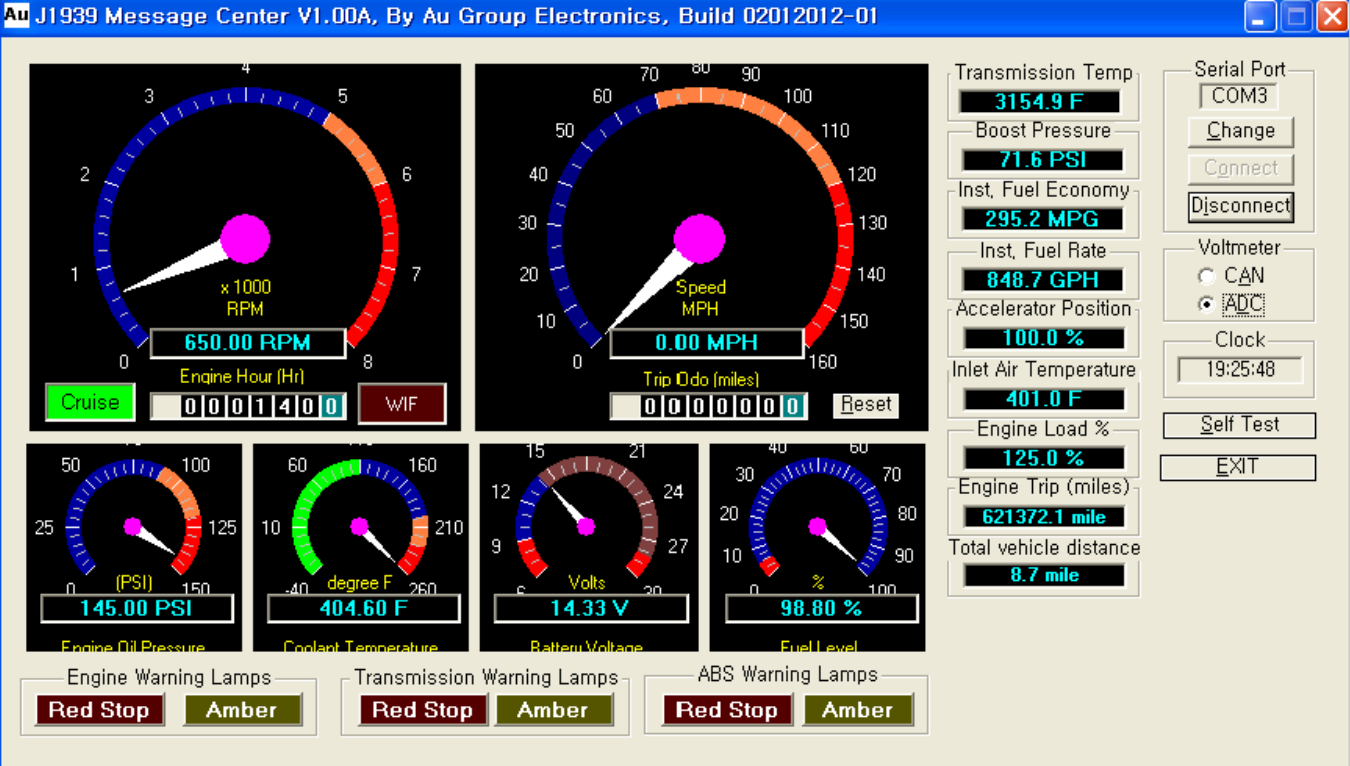}
    \caption{Au SAE J1939 message center 1.00A}
  \end{subfigure}%
  \caption{Graphic user interface for SAE J1939 network. This screenshot shows the tool's interface used for monitoring.}
  \label{fig:simulator_gui}
\end{figure}

The experimental setup is depicted in \Cref{fig:testbed}, illustrating the testbed constructed for the purpose of this study.
Laptop 1 has installed a GUI to monitor the simulator's signal values, as shown in \Cref{fig:simulator_gui}. 
Laptop 2 is utilized for CAN bus monitoring and to perform the role of an attacker. The CAN bus is monitored using a Kvaser USBcan Pro 2xHS v2. Python scripts are developed to inject the attacks, utilizing the CANlib provided by Kvaser.
For clarity in the experimental procedure, two laptops are used; however, it should be noted that the same operations can be conducted using a single laptop without affecting the outcome.
Because, in this study, we predicate on the assumption that an attacker will gain unauthorized access to the SAE J1939-based in-vehicle network. 
Such access can be obtained from the on-board diagnostics (OBD)-II port through physical intrusion or in-vehicle telematics such as infotainment systems \cite{jeong2023XCANIDS}. 
Additionally, attackers may exploit a parked vehicle by leveraging a CAN dongle tapped into an exposed rear camera.
Both the Au SAE J1939 Simulators (Generation II) and the Au SAE J1939 MCS are connected to the CAN bus, which facilitates packet transmission and reception. 
The CAN bus configuration for each simulator necessitates a 12V power supply.

\section{Experiment} \label{sec:experiment}

\begin{figure}[!t]
  \centering
  \begin{subfigure}{.33\textwidth}
    \centering
    \captionsetup{justification=centering}
    \includegraphics[width=0.9\linewidth]{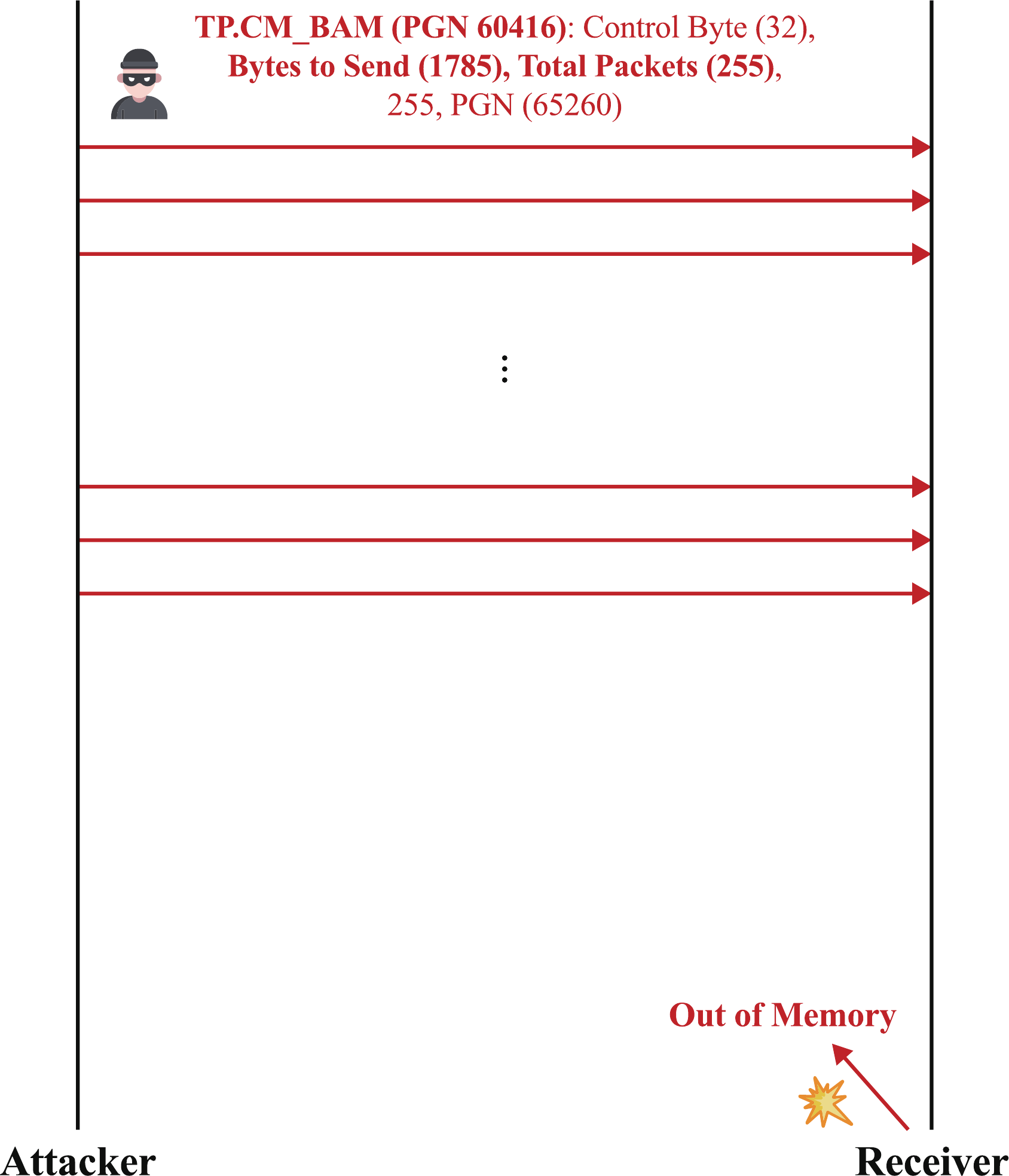}
    \caption{Out of memory using\\TP.CM\_BAM attack}
    \label{fig:scenario_attack11}
  \end{subfigure}%
  \begin{subfigure}{.33\textwidth}
    \centering
    \captionsetup{justification=centering}
    \includegraphics[width=0.9\linewidth]{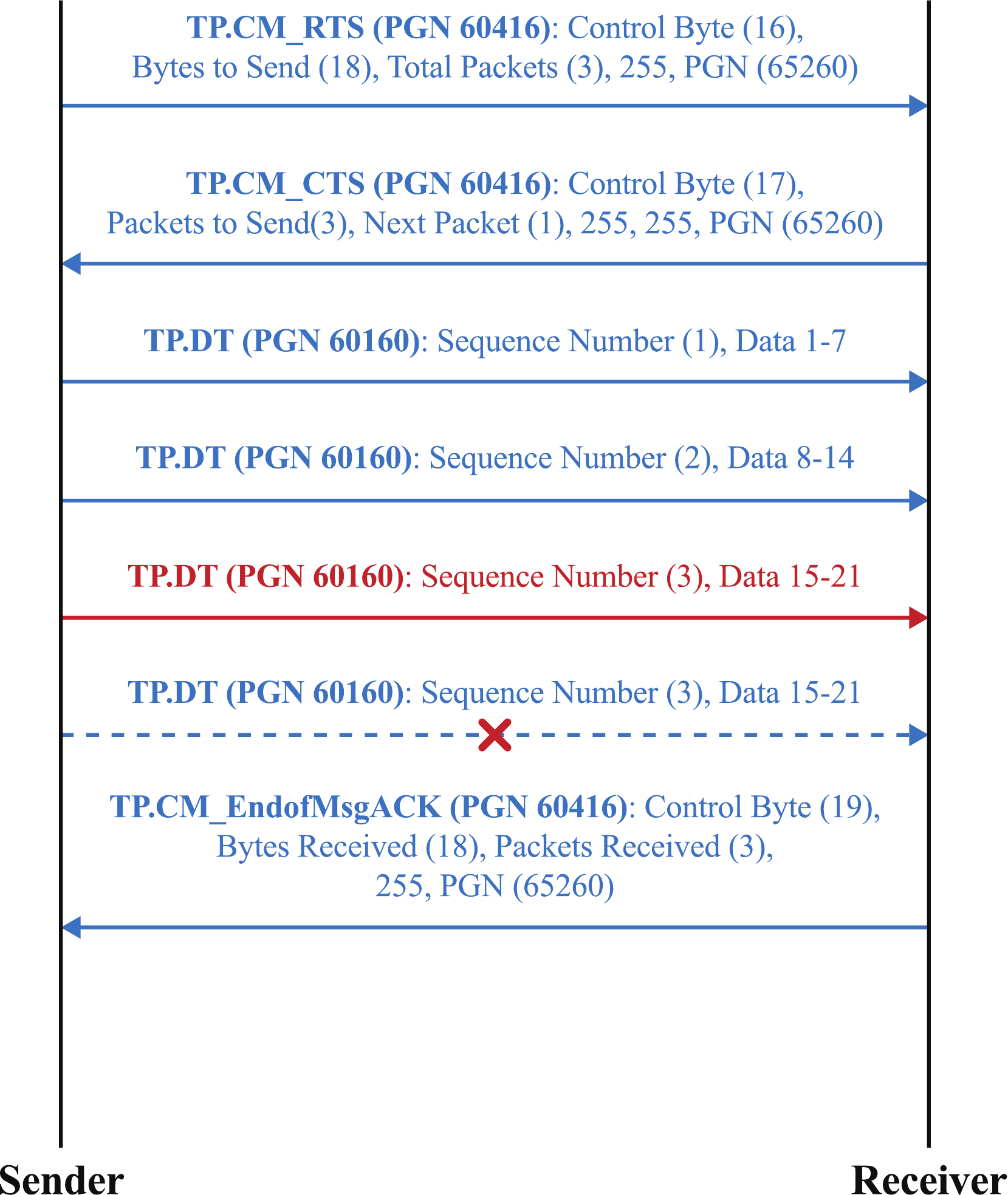}
    \caption{Overwrite memory using\\TP.DT attack}
    \label{fig:scenario_attack13}
  \end{subfigure}%
  \begin{subfigure}{.33\textwidth}
    \centering
    \captionsetup{justification=centering}
    \includegraphics[width=0.9\linewidth]{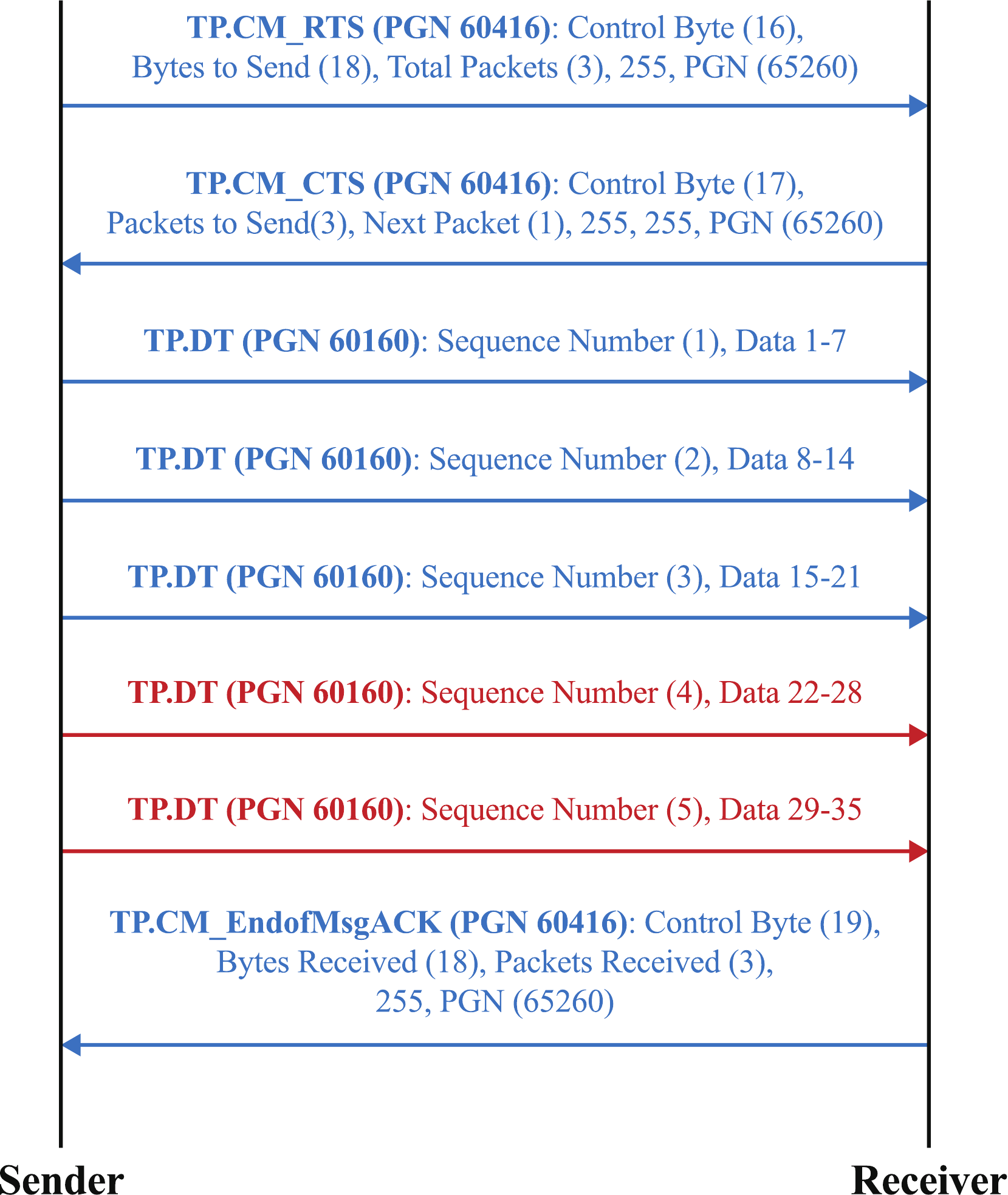}
    \caption{TP.DT spoofing attack\\{$\ $}}
    \label{fig:scenario_attack14}
  \end{subfigure}%
  \vspace{5mm}

  \caption{Selected attack scenarios in SAE J1939. They are showcasing 3 of 14 scenarios with notable impact. Comprehensive details on all discovered attack scenarios are shown in the appendix.}
  \label{fig:attack_scenarios}
\end{figure}

In this section, we describe the attack scenarios dedicated to SAE J1939, excluding attack scenarios commonly working in the CAN bus (e.g., fuzzing, DoS, spoofing, replay, suspension, and masquerade attacks).
Specifically, we only consider the scenarios involving the Request and TP.
We exclude scenarios involving Request2, as they are defined by the manufacturer.
We develop 14 attack scenarios and conduct experiments in a testbed environment.
Among them, seven scenarios have been previously proposed in existing research, while we introduce other seven scenarios as new proposals.
This chapter describes each attack scenario and its results.
In \Cref{fig:attack_scenarios} and \Cref{fig:appendix_attack_scenarios}, messages from normal ECUs are shown in blue, while messages from attackers and those generated by attacks are in red.
Additionally, normal messages that should have occurred but did not are represented by dashed lines.
\Cref{fig:results_all_attacks} shows the results for each attack.
During all attacks, injections are made such that the testbed baud rate utilized only 20-50\% of the maximum capacity of messages, not 100\%.
During normal operation, approximately 20\% of the baud rate is utilized.
This controlled injection rate is used to avoid the typical DoS attacks observed in CAN, where messages are injected at a high rate with elevated priority AIDs, pushing normal messages out of the bus due to baud rate saturation.
By injecting fewer messages, we can confirm the effects are due to the newly proposed SAE J1939 attack scenarios and not pre-existing ones.
We do not alter priority when injecting attacks, retaining the priority used in the existing testbed.
Since changing SAE J1939 priority will alter the AID, which could be easily detected.
Even if viable, such attacks are excluded from our study due to their low risk.
All tests are conducted in a black-box environment without access to the source code or debug information.
Furthermore, as actual ECUs are not in operation, no information is exposed, even in a buffer overflow.

\subsection{Request Overload Attack}
\subsubsection{Attack Scenario and Testing}
The first attack scenario is a DoS attack, as proposed by Mukherjee \textit{et al.}\cite{mukherjee2016practical}, where the attacker paralyzes the target ECU using a flood of Request messages.
\Cref{fig:scenario_attack1} illustrates the Request overload attack.
To determine if a PGN is currently in use, a message is sent using PGN 59904.
In the testbed, specific ECUs periodically send Request messages inquiring about the use of PGNs 65227, 65253, 65254, and 65260.
We target PGN 65263 for our attack, which periodically transmits engine fluid information, as the target ECU.
This PGN is not originally receiving Request messages in the testbed.
It is expected that PGN 65263, receiving a flood of Request messages, will be unable to provide normal services.

\subsubsection{Results}
During normal operation, PGN 65263 transmits messages at an average interval of 0.5 seconds.
Upon the injection of the attack, the simulator displays engine fluid information with a reading of 0, indicating a failure to provide accurate information.
\Cref{fig:results_attack1} shows that the message generation interval for PGN 65263 changed and is between 0.27 seconds and 0.72 seconds upon the attack injection.
The higher the injection speed, the more pronounced the effects of the attack.
This indicates that the effects are due to the attack.
Injecting the attack at very high speeds to occupy most of the baud rate also results in a reduced frequency of PGN 65263 messages, similar to findings by previous researchers.
However, as this proposes an attack scenario difficult to detect by intrusion detection models, the message delay is considered an effect of the attack.
The capability to target specific ECUs and the potential for significant damage even with such delays, especially at high speeds, suggests an effective attack scenario.

\subsection{Malicious Acknowledgment Attack}
\subsubsection{Attack Scenario and Testing}
The second attack scenario introduces a new method where an attacker manipulates and sends an Acknowledgment for a Request message. This scenario aims to exploit communication process vulnerabilities by issuing Acknowledgment messages that may not match authentic Request messages, potentially leading to unauthorized activities or data breaches.
\Cref{fig:scenario_attack2} illustrates this malicious Acknowledgment attack, showing how an attacker can compromise the communication protocol's integrity and possibly disrupt or weaken the security of the networked system.
In SAE J1939, Request messages are utilized to identify active PGNs.
Such information is crucial for directing messages to specific PGNs and ensuring that other ECUs do not use duplicate PGNs.
Consequently, if a PGN is reported as inactive or unresponsive when in use, it can halt message transmission to that PGN or lead to its use by another ECU, thereby causing network confusion.
Our experiment focuses on PGN 65263, which typically receives Request messages under normal operation.
The goal is to dispatch false responses more swiftly than the actual response messages.
For testing, PGN 65263 must respond with 0 (ACK) in the Acknowledgment message through PGN 59392 to indicate that it is in existing use, but this is changed to 1 (NACK), 2 (Access Denied), and 3 (Cannot Respond) before being sent.
Misrepresenting an active PGN as inactive is expected to prevent the ECU from receiving previously accessible messages.

\subsubsection{Results}
Unfortunately, PGN 59392 is not utilized in our testbed, making the testing process challenging.
We attempted to forcefully insert the values 1 (NACK), 2 (access denied), and 3 (unable to respond) into the Acknowledgment message for the Request message sent to PGN 65263, but this had no noticeable effect.
This is expected because our testbed does not consider PGN 59392, so there may be no response.
However, when only PGNs that return an Acknowledgment of 0 (ACK) in response to periodic Request messages are communicated with, even minimal packet injection can constitute a considerable attack threat.

\subsection{Malicious TP.CM\_RTS Attack}
\subsubsection{Attack Scenario and Testing}
The third attack scenario, similar to the scenario proposed by Mukherjee \textit{et al.}\cite{mukherjee2016practical}, involves the attacker injecting TP.CM\_RTS messages to alter the number of packets or the length of data to be transmitted.
\Cref{fig:scenario_attack3} illustrates the malicious TP.CM\_RTS attack.
The attacker exploits the fact that when multiple TP.CM\_RTS messages are injected from the same SA, the most recent message is accepted, and by timing the injections, the attacker sends TP.CM\_RTS messages significantly alter the number of packets a normal ECU would send, either reducing or increasing the total number.
In SAE J1939, when a normal ECU needs to transmit a message exceeding eight bytes, it informs the number of packets it will send within TP.CM\_RTS to another ECU.
When establishing a connection, the receiving ECU verifies the number and length of incoming packets, accordingly reserving the necessary buffer space.
In the testbed, During normal operation, ECU 0 sends TP.CM\_RTS message to ECU 249 to transmit multi-packets.
At this moment, it is declared that a total of three packets will be sent.
ECU 249 is required to process the information from ECU 0's TP.CM\_RTS message and respond with TP.CM\_CTS to facilitate the transmission of three messages. However, an attacker targeting the same PGN may transmit a malicious TP.CM\_RTS immediately following the legitimate message, altering the indication to send two to four messages instead. Subsequently, ECU 249 reacts by issuing TP.CM\_CTS to transmit the number of packets as specified by the altered TP.CM\_RTS.
Suppose an attacker sends a message that reduces the number of packets a normal ECU tries to send. In that case, the receiving ECU will believe that all messages have been received, even though fewer messages were sent, and will prematurely terminate the session using TP.CM\_EndofMsgACK.
As a result, it is anticipated that packets from the normal transmitting ECU will go undelivered.
On the other hand, if the attacker issues a message to increase the packet count beyond what the normal transmitting ECU plans, the receiving ECU will issue a TP.CM\_CTS message to request additional packets, potentially leading to data leakage.
When sending the attack message, consideration is given to both altering only the number of packets while keeping the message length the same and changing the message length to match the altered packet count.

\begin{table*}[t]
\setlength{\tabcolsep}{3pt}
\centering
\caption{Test cases of TP.CM\_RTS spoofing attacks.}
\label{table:tp_cm_rts_spoofing}
\begin{adjustbox}{max width=\linewidth}
\begin{tabular}{rl|rrrrr}
\hline
\multicolumn{2}{l|}{\textbf{Type of Message}}                                       & \textbf{Normal} & \textbf{Case 1} & \textbf{Case 2} & \textbf{Case 3} & \textbf{Case 4} \\ \hline
\multicolumn{1}{l|}{\multirow{2}{*}{TP.CM\_RTS}}         & Bytes to send   & 18     & 18          & 18          & \textbf{10}          & \textbf{22}          \\ 
\multicolumn{1}{l|}{}                                    & Total packets   & 3      & \textbf{2}           & \textbf{4}           & \textbf{2}           & \textbf{4}           \\ \hline
\multicolumn{1}{l|}{\multirow{2}{*}{TP.CM\_CTS}}         & Packets to send   & 18     & 18          & 18          & \textbf{10}          & \textbf{22}          \\ 
\multicolumn{1}{l|}{}                                    & Next packet   & 3      & \textbf{2}           & \textbf{4}           & \textbf{2}           & \textbf{4}           \\ \hline
\multicolumn{2}{l|}{Number of TP.DTs transmitted}                                       & 3      & \textbf{2}           & 3           & \textbf{2}           & 3           \\ \hline
\multicolumn{1}{c|}{\multirow{2}{*}{TP.CM\_EndofMsgACK}} & Bytes received  & 18     & 18          & -           & \textbf{10}          & -           \\ 
\multicolumn{1}{c|}{}                                    & Packets received & 3      & \textbf{2}           & -           & \textbf{2}           & -           \\ \hline
\end{tabular}
\end{adjustbox}
\end{table*}

\subsubsection{Results}
During normal operations, the transmitting and receiving ECUs synchronize on the number of packets to be sent, guaranteeing a smooth exchange and preventing packet loss.
However, the results after the injection of the attack are shown in Table \ref{table:tp_cm_rts_spoofing}.
Regardless of the change in packet size, if the attacker transmits a TP.CM\_RTS message specifying a reduced number of packets, only two packets are sent instead of the expected three, leading to a loss of information.
If the attacker modifies the number of packets to larger quantities, transmitting ECU sends three TP.DT packets and then send a TP.Conn\_Abort message to forcibly close the session.
This consequence likely originates from the programming of ECU 0, which limits it to sending a maximum of three packets and triggers a TP.Conn\_Abort message in response to any unexpected messages.
This consequence is expected from the programming of ECU 249, which limits it to sending a maximum of three packets and triggers a TP.Conn\_Abort message in response to any unexpected messages.
These observations indicate that the effectiveness of the attack is maintained, irrespective of whether the manipulated number of packets is higher or lower than anticipated.

\subsection{Connection Exhaustion Attack}
\subsubsection{Attack Scenario and Testing}
The fourth attack scenario, as proposed by Mukherjee \textit{et al.}\cite{mukherjee2016practical}, involves the attacker periodically sending TP.CM\_CTS messages to maintain a session.
\Cref{fig:scenario_attack4} illustrates the TP.CM\_CTS DoS attack.
While a normal ECU needs to establish a session with the receiving ECU to send multi-packet messages, the attacker continually sends messages to occupy the session.
Only one session can be established at a time.
In SAE J1939 protocol, a session is requested by sending TP.CM\_RTS, and subsequently established and maintained with TP.CM\_CTS to facilitate the reception of multi-packet messages.
A TP.CM\_CTS message sent within 1250 milliseconds after data transmission indicates either a failure to receive a specific multi-packet or a request for its retransmission, thus ensuring the continuation of the session.
Consequently, if an attacker occupies a session intended to receive multi-packet messages, the legitimate information fails to be received.
In the testbed, ECUs 0 and 249 usually establish a session and exchange information at 45 to 60 seconds intervals.
When sustaining a session through a packet retransmission request, the request must be sent before the arrival of the next TP.DT message to maintain the session. Therefore, the attack period is set within a 50-millisecond window.
During the attack, since additional sessions for multi-packet transmission and reception cannot be initiated, it is anticipated that ECU 249 will cease to receive the messages it normally receives periodically.

\subsubsection{Results}
Upon the execution of the attack, ECU 0 and ECU 249 are observed to be unable to initiate a new session, resulting in the cessation of their regular information exchanges.
If a significant volume of multi-packets is transmitted and received within the testbed, the time until the complete cessation of the session could be prolonged. It presents the opportunity to carry out attacks over extended intervals lasting less than 1250 milliseconds.
\Cref{fig:results_attack4} illustrates the modified status of multi-packet transmissions before and after the execution of the attack.
After the attack, the targeted ECU is prevented from receiving the multi-packet messages that are normally transmitted and received.
The resumption of normal multi-packet transmissions after the attack indicates a temporary disruption, thus confirming the effectiveness of the attack.

\subsection{TP.CM\_BAM Block Attack}
\subsubsection{Attack Scenario and Testing}
The fifth attack scenario, introduced by Chatterjee \textit{et al.}\cite{chatterjee2023exploiting}, extends the principles of the fourth attack scenario specifically to target the ECUs responsible for broadcasting messages.
\Cref{fig:scenario_attack5} demonstrates the TP.CM\_BAM block attack, which obstructs the transmission of TP.CM\_BAM messages that were previously sent periodically by occupying the session.
It is identical to the fourth attack scenario, but in this case, the target ECU is ECU 0, which is responsible for transmitting the TP.CM\_BAM message instead of ECU 249.
To interrupt the transmission of TP.CM\_BAM messages by ECU 0, the attacker seizes the communication session using TP.CM\_RTS and TP.CM\_CTS messages.
Consequently, ECU 0 is expected to be incapacitated and unable to continue sending TP.CM\_BAM messages.

\subsubsection{Results}
Under normal operations, ECU 0 normally broadcasts TP.CM\_BAM messages at intervals of approximately 2.5 seconds.
\Cref{fig:results_attack5} illustrates the change in the frequency of TP.CM\_BAM message broadcasting before and after the attack is executed.
Observations following the attack reveal that the target ECU cannot send TP.CM\_BAM messages.
The return to regular TP.CM\_BAM message transmissions after the attack validate the temporary disruption caused by the attack, underscoring its effectiveness.

\subsection{Memory Leak using TP.CM\_CTS Attack 1}
\subsubsection{Attack Scenario and Testing}
The sixth attack scenario, proposed by Chatterjee \textit{et al.}\cite{chatterjee2023exploiting}, involves the attacker sending malicious TP.CM\_CTS messages to cause a data leakage.
This method of attack manipulates the mechanism used by the communication protocol to manage multi-packet transmissions, exploiting vulnerabilities within the system. 
\Cref{fig:scenario_attack6} represents this approach, illustrating a memory leak instance through a TP.CM\_CTS attack, underscoring the protocol's vulnerability to compromise data integrity and system stability.
In SAE J1939, TP.CM\_RTS is utilized to signal the intention to transmit $N$ multi-packets.
Following this, the ECU received the TP.CM\_RTS responds with a TP.CM\_CTS message, indicating readiness to receive packets from $1$ to $N$.
However, an attacker can exploit this process by sending a malicious TP.CM\_CTS message that requests packets starting from a sequence number greater than or equal to $N+1$, thus exceeding the original packet range and potentially causing a data leak.
ECU 0 initiates a session with ECU 249 via TP.CM\_RTS to send three multi-packets in the testbed.
ECU 249, intending to comply, responds with a TP.CM\_CTS message aimed at receiving three packets, starting with packet 1.
However, the scenario is compromised by an attacker who issues a TP.CM\_CTS message requesting three packets starting from packet 5, a sequence beyond the agreed-upon transmission range.
This action increases the likelihood of data leakage, especially if the targeted ECU lacks safeguards to reject or handle such invalid TP.CM\_CTS messages appropriately.

\subsubsection{Results}
Table \ref{table:tp_cm_cts_bof} illustrates the sequence of packet occurrences before and after the attack.
After the attack is launched, ECU 249 instructs ECU 0 to transmit three packets, starting with the fifth packet, resulting in the system skipping the first and second TP.DT packets and immediately send the third TP.DT message.
Following this, a TP.CM\_CTS message is transmitted, requesting the transmission of zero packets starting with the fourth packet, without any subsequent session termination message.
If ECU 0 is to send the normal TP.CM\_RTS and ECU 249 is to respond with TP.CM\_CTS during this time, the session would abruptly end with a TP.Conn\_Abort message.
This result is supposed by the limitation of the testbed, indicating a significant risk of data leakage should such an attack be executed against actual ECUs or trucks.
ECU 249 skips the sequence of packets it is originally scheduled to send and attempts to send from a higher number, confirming the effectiveness of this type of attack.

\begin{table*}[t]
\setlength{\tabcolsep}{10pt}
\centering
\caption{Results of memory leak using TP.CM\_CTS attack 1.}
\label{table:tp_cm_cts_bof}
\begin{adjustbox}{max width=\linewidth}
\begin{tabular}{r|ll|rr}
\hline
\textbf{Message sequence}    & \multicolumn{2}{l|}{\textbf{Type of message}}                                                                  & \textbf{Normal}                                          & \textbf{Attack}                                          \\ \hline
\multirow{2}{*}{1} & \multicolumn{1}{l|}{\multirow{2}{*}{TP.CM\_RTS}}                      & Bytes to send                 & 18                                              & 18                                              \\ 
                   & \multicolumn{1}{l|}{}                                                 & Total packets                 & 3                                               & 3                                               \\ \hline
\multirow{2}{*}{2} & \multicolumn{1}{l|}{\multirow{2}{*}{TP.CM\_CTS}}                      & Packets to send               & 3                                               & 3                                               \\ 
                   & \multicolumn{1}{l|}{}                                                 & Next packet                   & 1                                               & \textbf{5}                                               \\ \hline
3                  & \multicolumn{2}{l|}{\begin{tabular}[c]{@{}l@{}}Number of TP.DTs transmitted \\ (last sequence number)\end{tabular}} & \begin{tabular}[c]{@{}r@{}}3\\ (3)\end{tabular} & \begin{tabular}[c]{@{}r@{}}1\\ (3)\end{tabular} \\ \hline
\multirow{2}{*}{4} & \multicolumn{1}{l|}{\multirow{2}{*}{TP.CM\_EndofMsgACK}}              & Bytes received                & 18                                              & -                                               \\ 
                   & \multicolumn{1}{l|}{}                                                 & Packets received              & 3                                               & -                                               \\ \hline                 \multirow{2}{*}{5} & \multicolumn{1}{l|}{\multirow{2}{*}{TP.CM\_CTS}}                                       & Packets to send               & -                                               & \textbf{0}                                               \\ 
                   & \multicolumn{1}{l|}{}                                                 & Next packet                   & -                                               & \textbf{4}                                               \\ \hline
\end{tabular}
\end{adjustbox}
\end{table*}

\subsection{Memory Leak using TP.CM\_CTS Attack 2}
\subsubsection{Attack Scenario and Testing}
The seventh attack scenario, as proposed by Chatterjee \textit{et al.}\cite{chatterjee2023exploiting}, involves an attacker sending fake TP.CM\_CTS messages to induce data leakage.
This scenario shares similarities with the sixth attack scenario but differs in approach. Instead of requesting packets starting with a higher sequential number than originally intended, the attacker solicits more packets than initially agreed upon, resulting in data leakage.
\Cref{fig:scenario_attack7} shows the memory leak during TP.CM\_CTS attack 2. Despite the normal ECU's intention to transmit three packets starting with number 1, the attacker's request for ten packets from the same starting point leads to data leakage.
ECU 0 initiates a session connection request with ECU 249 via TP.CM\_RTS to transmit three multi-packets in the testbed.
Following this, ECU 249 aims to dispatch three packets via a TP.CM\_CTS message, starting with the first packet.
The attacker intervenes by disseminating a malicious TP.CM\_CTS message that erroneously requests the transmission of ten packets, starting with the first packet.
This scenario poses a risk of data leakage if the ECU lacks the programming to handle such invalid TP.CM\_CTS messages effectively.

\subsubsection{Results}
The outcomes of the attack are the same as those observed in the memory leak during TP.CM\_CTS attack 1.
This result is also supposed by the limitation of the testbed.
The continued attempt by ECU 249 to transmit additional packets proves the attack's effectiveness.

\subsection{Memory Leak using TP.CM\_CTS Attack 3}
\subsubsection{Attack Scenario and Testing}
The eighth attack scenario presents an unexplored approach where an attacker induces memory leakage in the target ECU by sending a TP.CM\_CTS message to initiate packet transmission, even when there is no TP.CM\_RTS message originally sent by the target. This attack leads the targeted ECU to prepare to receive data that is not legitimately requested, causing potential memory resource issues.
\Cref{fig:scenario_attack8} illustrates TP.CM\_CTS attack 3.
The attacker exploits this by issuing TP.CM\_CTS messages to request the transmission of multi-packets, aiming to cause a memory leakage without a preceding TP.CM\_RTS message.
Within the testbed, if ECU 0 has not dispatched a TP.CM\_RTS message, the attacker proceeds to send a TP.CM\_CTS message to instigate multi-packet transmission.
Consequently, upon receiving such a message, the target ECU is expected to suffer from memory leakage.

\subsubsection{Results}
Despite the injection of the attack, no impact is observed.
The injected messages include both TP.CM\_CTS messages previously encountered in the testbed and new packets that have not been observed before.
Ultimately, it was determined that in the absence of a TP.CM\_RTS message, the arrival of a TP.CM\_CTS message did not elicit a response.
 
\subsection{TP.CM\_EndofMsgACK Interruptions Attack}
\subsubsection{Attack Scenario and Testing}
The ninth attack scenario introduces an unknown disruption technique, terminating the session by injecting a TP.CM\_EndofMsgACK message.
\Cref{fig:scenario_attack9} shows this newly proposed TP.CM\_EndOfMsgACK interruption attack.
When ECUs have established a session for transmitting and receiving multi-packets and are ready to send data, the attacker intervenes by sending a TP.CM\_EndOfMsgACK message, forcibly terminating the session.
The attacker exploits the forced termination of the session during multi-packet transmission, thereby interrupting the normal data exchange.
ECU 0 initiates a session with ECU 249 using the TP.CM\_RTS message for multi-packet transmission in the testbed.
The attacker disrupts the session during this process by injecting a malicious TP.CM\_EndofMsgACK message.
This attack is expected to cause the premature termination of the session.

\subsubsection{Results}
While injecting the attack, ECU 0 discovers that any attempt to initiate a session is promptly terminated, thus preventing the transmission of essential information.
\Cref{fig:results_attack9} shows the impact on multi-packet transmission before and after the attack. 
Following the attack's introduction, ECU 0 cannot continue transmitting multi-packet messages because the session is forcibly terminated. 
We can confirm the attack is successful because there are no packet exchanges between ECU 0 and ECU 249 during the attacks, and we observe the packet exchanges are normal after the attacks have stopped.

\subsection{TP.Conn\_Abort Interruptions Attack}
\subsubsection{Attack Scenario and Testing}
The 10th attack scenario, as proposed by Murvay \textit{et al.}\cite{murvay2018security}, outlines a DoS attack that leverages the TP.Conn\_Abort command to forcibly terminate a connection. This method disrupts normal communication sessions between ECUs within vehicular networks.
In \Cref{fig:scenario_attack10}, this scenario depicts the TP.Conn\_Abort interruptions attack in action. During a legitimate communication session, where a normal ECU attempts to connect and send messages to another ECU, the attacker intervenes by issuing a TP.Conn\_Abort command. This command effectively and prematurely ends the session, preventing the successful transmission of messages.
This attack exploits SAE J1939 protocol’s provision for session termination, using it maliciously to disrupt normal operations and service availability, thereby demonstrating a significant vulnerability in the protocol’s design concerning security against DoS attacks.
SAE J1939 standard includes a protocol, TP.Conn\_Abort, designed to terminate a connection for a variety of reasons forcefully.
Attackers can exploit this feature to disrupt the normal communication between ECUs by forcibly ending sessions intended for multi-packet transmission.
ECU 0 initiates a session connection to ECU 249 using TP.CM\_RTS for the purpose of sending multi-packets in the testbed.
During this process, an attacker intervenes by dispatching a malicious TP.Conn\_Abort message, effectively terminating the session prematurely.
SAE J1939 protocol reserves up to 255 different reasons for connection termination, and in this experiment, each of these reasons was tested to understand the impact of such attacks on normal ECU communication, excluding those that are manufacturer-defined and intended for specific use.
The expectation after an attack is the forceful termination of the session dedicated to the sending and receiving of multi-packets.

\subsubsection{Results}
ECU 0 is compelled to terminate its ongoing session upon introducing the attack, resulting in its inability to transmit the requisite information.
\Cref{fig:results_attack10} shows the impact on the existing multi-packet transmission status before and after the attack, highlighting the disruption caused.
Following the attack's injection, it is obvious that the ECU is forced to end the session, rendering it incapable of continuing its multi-packet message transmissions.
After the cessation of the attack, both ECU 0 and ECU 249 continued their exchange of multi-packet messages, demonstrating the effectiveness of the attack in disrupting the normal communication processes.

\subsection{Out of Memory using TP.CM\_BAM Attack}
\subsubsection{Attack Scenario and Testing}
The 11th attack scenario introduces an untried approach wherein the attacker dispatches an extensive volume of multi-packets to the target ECU using the TP.CM\_BAM protocol. 
This scenario is intended to overwhelm the target ECU by compelling it to allocate substantial resources in preparation for the receipt of the announced multi-packet transmission.
The primary objective of this attack is to deplete the target ECU's resources to such an extent that it becomes incapable of maintaining its availability for normal service operations.
\Cref{fig:scenario_attack11} shows the out of memory using TP.CM\_BAM attack, illustrating the process by which the attacker notifies the target ECU of the impending large-scale multi-packet transmission through a TP.CM\_BAM message.
ECU 249 is engaged in the transmission and reception of multi-packet data in the testbed.
The attacker exploits by sending a TP.CM\_BAM message to the target, ECU 249, signaling the intention to transmit a multi-packet message comprising 1,785 bytes.
Upon receipt of this message, ECU 249 allocates the necessary resources in anticipation of the incoming data.
This allocation of resources to handle the specified multi-packet transmission can strain the ECU if its resources are limited, potentially hindering its ability to maintain normal operations and service delivery. 

\subsubsection{Results}
After the injection of the attack, ECU 249 is rendered incapable of performing its normal services. 
\Cref{fig:results_attack11} shows the flow of multi-packets to and from ECU 249 before and after the attack, illustrating a significant disruption in the communication. 
Post-attack, ECU 249 ceases to send critical TP messages, including TP.CM\_RTS, TP.CM\_CTS, TP.DT, TP.CM\_EndofMsgACK, and TP.CM\_BAM, which are previously transmitted on a regularly.
Following the termination of the attack, both ECU 0 and ECU 249 resumed their normal multi-packet message exchanges, reaffirming that the attack successfully interfered with the usual communication protocols.

\subsection{Out of Memory using TP.CM\_RTS Attack}
\subsubsection{Attack Scenario and Testing}
The 12th attack scenario introduces a novel method of attack during which the perpetrator uses a TP.CM\_RTS message to overwhelm the target ECU by compelling it to allocate excessive resources in preparation for receiving a large volume of multi-packet data.
This method is intended to deplete the target ECU's resources, thereby rendering it unable to provide its regular services and effectively compromising its availability.
This strategy resembles the eleventh attack scenario in that the goal is to disrupt the target ECU's functionality through resource exhaustion. However, it leverages the TP.CM\_RTS message as a means to trigger the excessive consumption of resources.
\Cref{fig:scenario_attack12} illustrates the out of memory using TP.CM\_RTS attack.
In this scenario, the attacker sends a TP.CM\_RTS message to the target ECU requesting the transfer of a significant amount of multi-packet data.
Upon receiving this message, the target ECU allocates the necessary resources to accommodate the expected data flow. 
This allocation is prompted by the reception of the TP.CM\_RTS message, potentially leading to the exhaustion of the ECU's resources, compromising its operational capabilities.
ECU 249 engages in the transmission and reception of multi-packet data in the testbed. 
The attacker initiates the attack by transmitting a TP.CM\_RTS message to the targeted ECU 249, signaling the intention to send a multi-packet message of 1,785 bytes to ECU 249.
Upon receipt of this message, ECU 249 allocates the necessary resources in anticipation of receiving the specified multi-packet message.
However, if ECU 249's resources are limited, the allocation for this incoming data is expected to significantly impair its ability to maintain normal service operations, potentially leading to service degradation or denial of service situations due to resource exhaustion.

\subsubsection{Results}
\begin{figure}
  \centering

  \begin{subfigure}{.33\textwidth}
    \centering
    \captionsetup{justification=centering}
    \includegraphics[width=\linewidth]{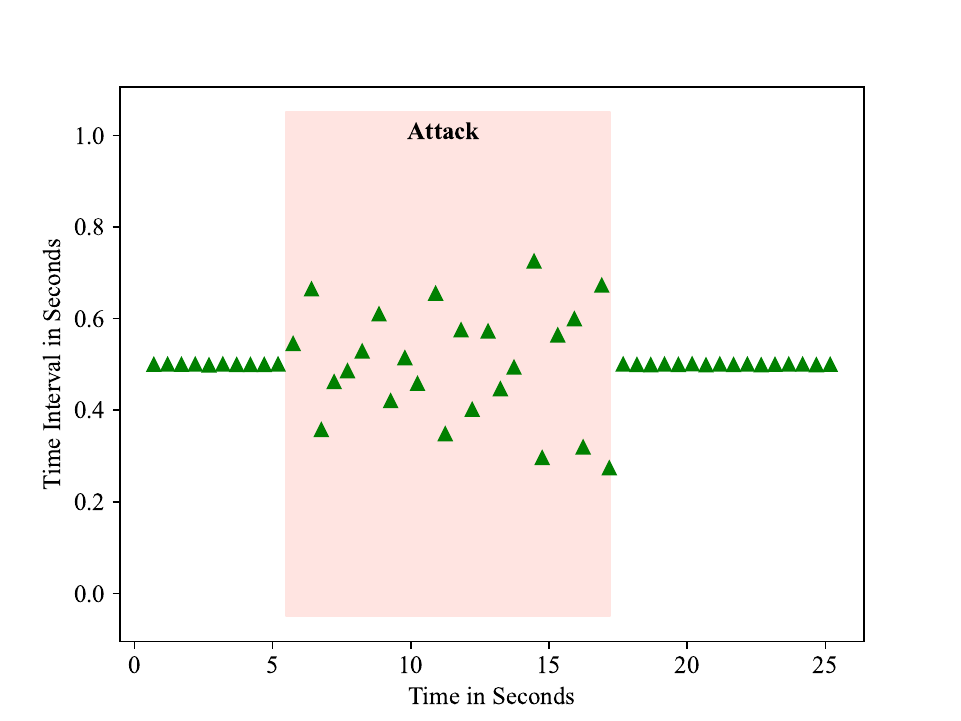}
    \caption{Result of the\\Request overload attack}
    \label{fig:results_attack1}
  \end{subfigure}%
  \begin{subfigure}{.33\textwidth}
    \centering
    \captionsetup{justification=centering}
    \includegraphics[width=\linewidth]{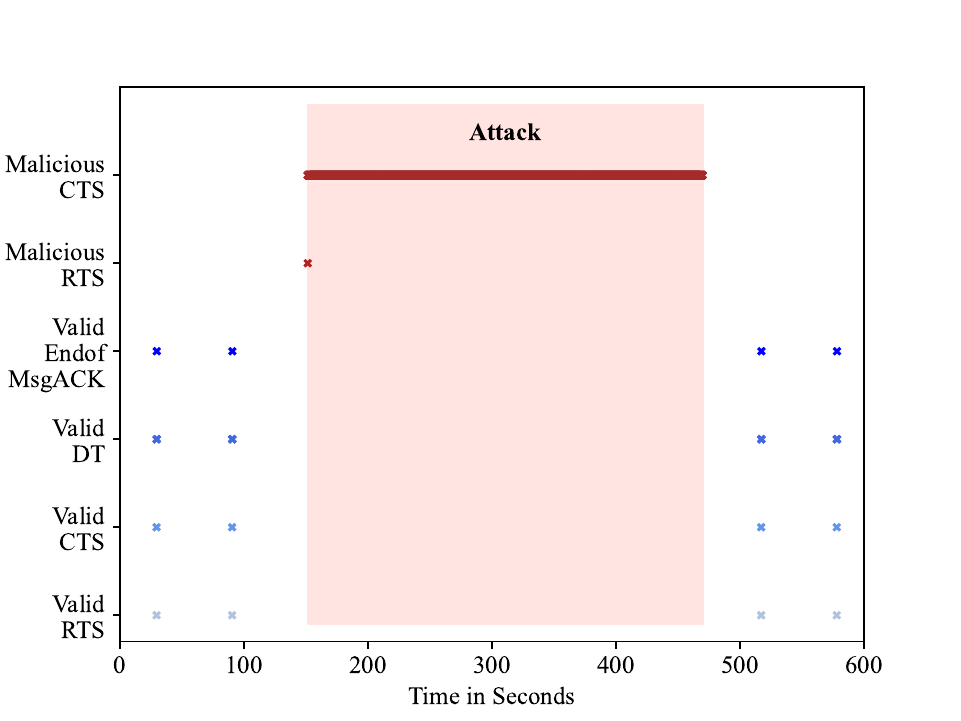}
    \caption{Result of the\\connection exhaustion attack}
    \label{fig:results_attack4}
  \end{subfigure}%
  \begin{subfigure}{.33\textwidth}
    \centering
    \captionsetup{justification=centering}
    \includegraphics[width=\linewidth]{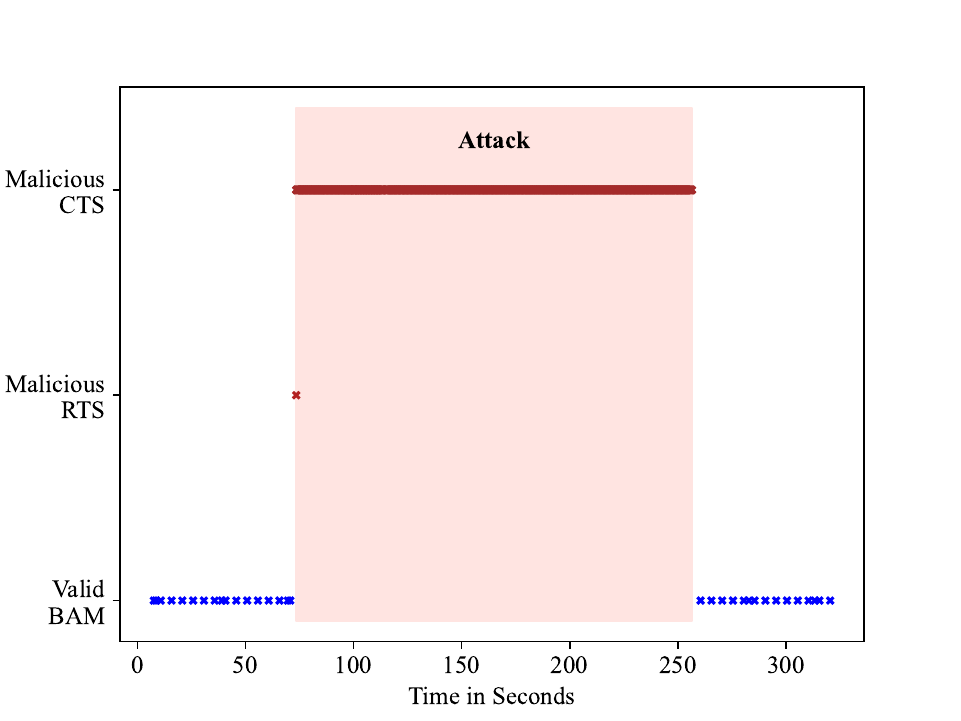}
    \caption{Result of the\\TP.CM\_BAM block attack}
    \label{fig:results_attack5}
  \end{subfigure}%
  
  \begin{subfigure}{.49\textwidth}
    \centering
    \captionsetup{justification=centering}
    \includegraphics[width=\linewidth]{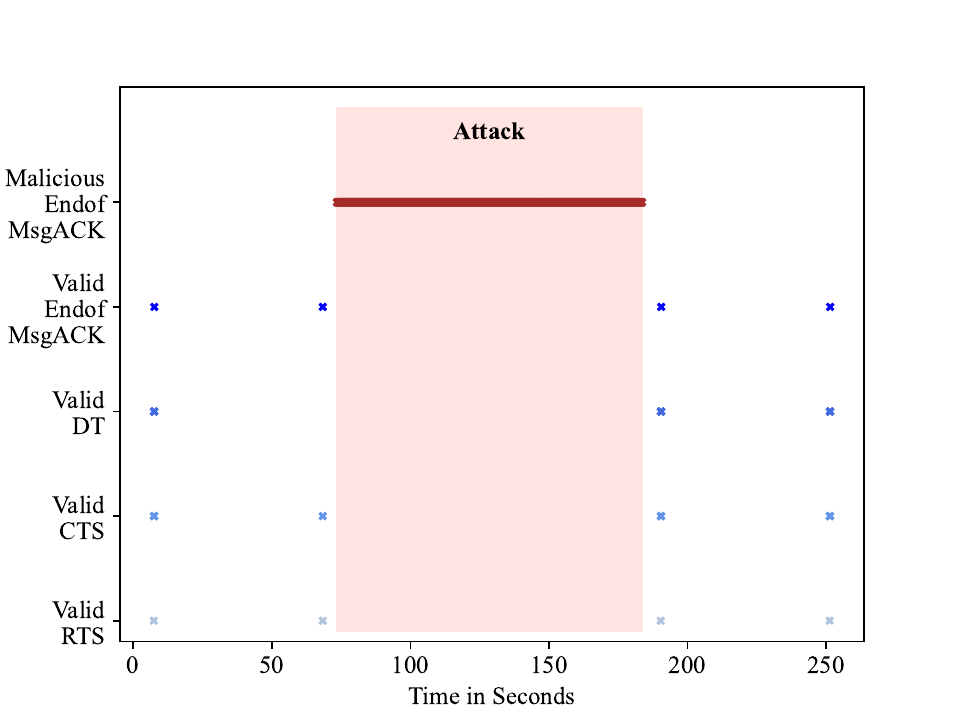}
    \caption{Result of the TP.CM\_EndofMsgACK\\interruptions attack}
    \label{fig:results_attack9}
  \end{subfigure}
  \begin{subfigure}{.49\textwidth}
    \centering
    \includegraphics[width=\linewidth]{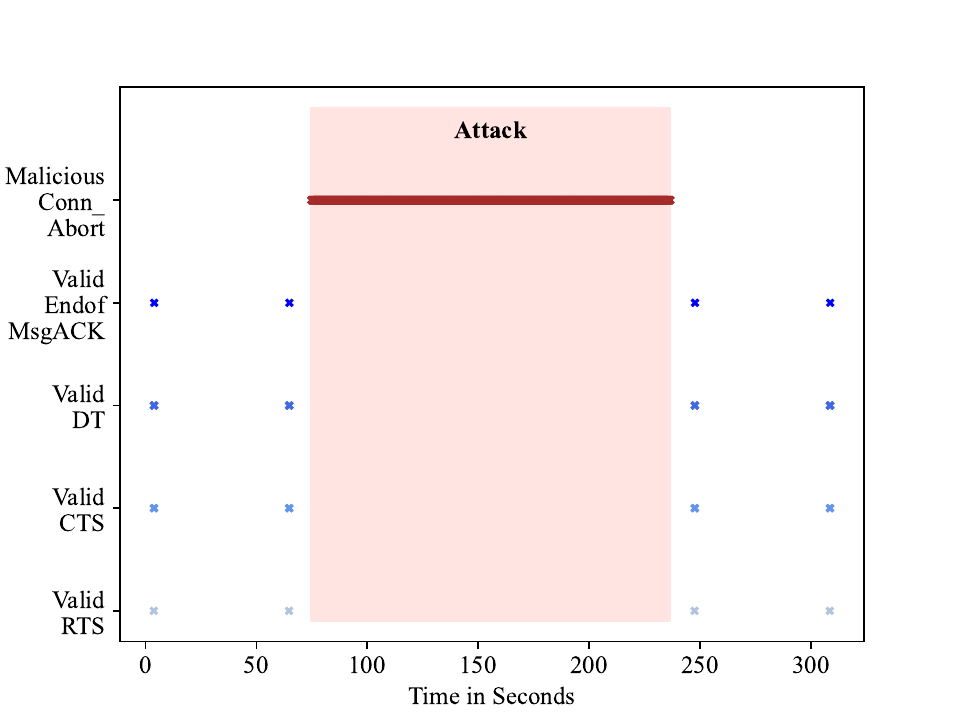}
    \caption{Result of the TP.Conn\_Abort\\interruptions attack}
    \label{fig:results_attack10}
  \end{subfigure}%
  
  \begin{subfigure}{.49\textwidth}
    \centering
    \captionsetup{justification=centering}
    \includegraphics[width=\linewidth]{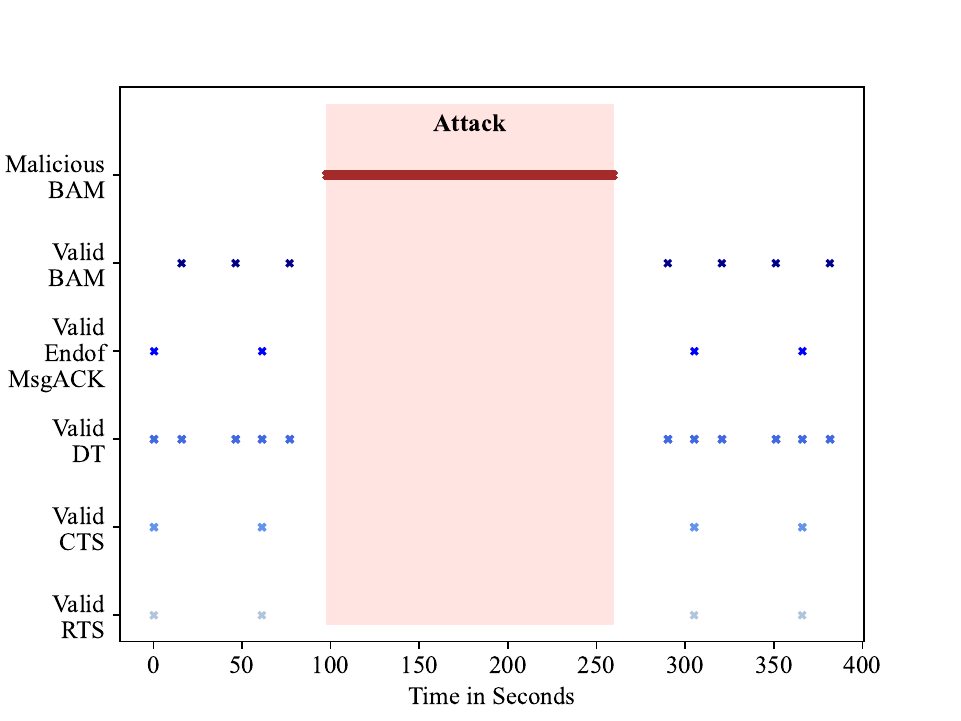}
    \caption{Result of the out of memory\\using TP.CM\_BAM attack}
    \label{fig:results_attack11}
  \end{subfigure}%
  \begin{subfigure}{.49\textwidth}
    \centering
    \captionsetup{justification=centering}
    \includegraphics[width=\linewidth]{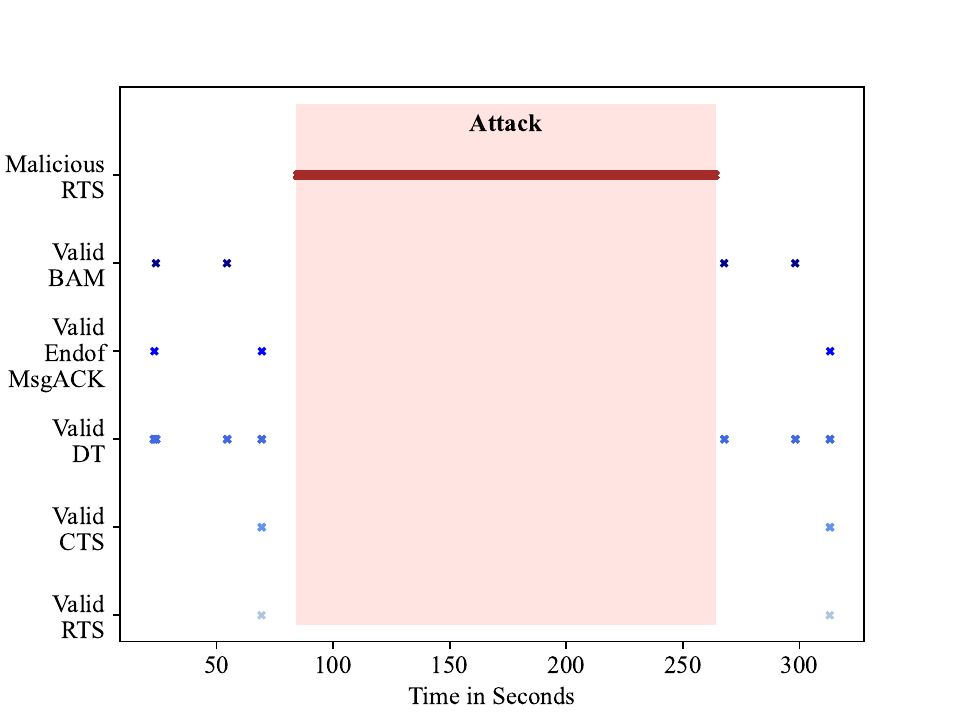}
    \caption{Result of the out of memory\\using TP.CM\_RTS attack}
    \label{fig:results_attack12}
  \end{subfigure}%

  \caption{Attack results for each scenario. The first sub-picture illustrates the immediate effect of an attack, where Request messages transition from periodic to irregular transmission, leading to unstable service. The subsequent six sub-pictures reveal a common consequence where, post-attack injection, previously normal message communications are abruptly halted.}
  \label{fig:results_all_attacks}
\end{figure}

Following the injection of the attack, akin to the out of memory using TP.CM\_BAM scenario, ECU 249 was incapacitated, losing its ability to perform normal services.
\Cref{fig:results_attack12} shows the multi-packet communications sent to and received by ECU 249 before and after the attack's execution. This provides clear evidence of how the attack impacts the data exchange with ECU 249, highlighting the change in the communication and possibly the inability of ECU 249 to maintain normal operations post-attack.

\subsection{Overwrite Memory using TP.DT Attack}
\subsubsection{Attack Scenario and Testing}
The 13th attack scenario introduces an unexplored method by which an attacker manipulates a specific segment of multi-packet data.
In \Cref{fig:scenario_attack13}, this technique, referred to as the overwrite memory using the TP.DT attack, shows how an attacker can exploit the timing of data injection.
During the normal operation of ECUs, it transmits and receives multi-packet information via TP.DT messages, the attacker strategically times the injection of malicious TP.DT messages.
This approach allows the attacker to overwrite existing data within the communication stream, potentially altering the intended message or compromising the integrity of the data being exchanged between ECUs.
ECU 0 initiates a connection for multi-packet transmission to ECU 249, utilizing TP.DT for message dispatch in the testbed. 
Three TP.DT messages are sent in rapid succession during this session. 
To compromise the data integrity of the second or third TP.DT message, an attack message is injected at a transmission before the normal TP.DT messages.

\subsubsection{Results}
Table \ref{table:overwrite_memory_tp_dt} presents the values of TP.DT messages during normal operation and when an attack is introduced.
Upon the injection of the attack, ECU 249 erroneously concludes that it has received all intended data, even though the received data does not originate from ECU 0. It terminates the session by issuing a TP.CM\_EndofMsgACK message.
This demonstrates the successful transmission of incorrect information to ECU 249 due to the attack.
Such an attack is particularly critical when vital information, such as information related to the engine or the speed of the vehicle, is transmitted in multi-packet formats, emphasizing the potential for a significant impact on the safety and operational integrity of the affected vehicle.

\begin{table*}[t]
\centering
\setlength{\tabcolsep}{5pt}
\caption{Malformed packet payloads used in TP.DT attack cases.}
\label{table:overwrite_memory_tp_dt}
\begin{adjustbox}{max width=\linewidth}
\begin{tabular}{r|ll|r|r|r}
\hline
\textbf{\makecell{Message\\sequence}} & \multicolumn{2}{l|}{\textbf{Type of message}}                                               & \textbf{Normal}                                                    & \textbf{Case 1}                                                    & Case 2                                                             \\ \hline
\multirow{2}{*}{1}       & \multicolumn{1}{l|}{\multirow{2}{*}{TP.CM\_RTS}}                 & Bytes to send            & 18                                                                 & 18                                                                 & 18                                                                 \\  
                         & \multicolumn{1}{l|}{}                                            & Total packets            & 3                                                                  & 3                                                                  & 3                                                                  \\ \hline
\multirow{2}{*}{2}       & \multicolumn{1}{l|}{\multirow{2}{*}{TP.CM\_CTS}}                 & Packets to send          & 3                                                                  & 3                                                                  & 3                                                                  \\ 
                         & \multicolumn{1}{l|}{}                                            & Next packet              & 1                                                                  & 1                                                                  & 1                                                                  \\ \hline
3                        & \multicolumn{2}{l|}{\begin{tabular}[c]{@{}l@{}}TP.DT data\\ (sequence number)\end{tabular}} & \begin{tabular}[c]{@{}r@{}}31 4D 38 47 44 4D 39\\ (1)\end{tabular} & \begin{tabular}[c]{@{}r@{}}31 4D 38 47 44 4D 39\\ (1)\end{tabular} & \begin{tabular}[c]{@{}r@{}}31 4D 38 47 44 4D 39\\ (1)\end{tabular} \\ \hline
4                        & \multicolumn{2}{l|}{\begin{tabular}[c]{@{}l@{}}TP.DT data\\ (sequence number)\end{tabular}} & \begin{tabular}[c]{@{}r@{}}41 38 4B 50 30 34 32\\ (2)\end{tabular} & \begin{tabular}[c]{@{}r@{}}41 38 4B 50 30 34 32\\ (2)\end{tabular} & \begin{tabular}[c]{@{}r@{}}\textcolor{red}{FF FF FF FF FF FF 00}\\ (2)\end{tabular} \\ \hline
5                        & \multicolumn{2}{l|}{\begin{tabular}[c]{@{}l@{}}TP.DT data\\ (sequence number)\end{tabular}} & \begin{tabular}[c]{@{}r@{}}30 30 30 2A FF FF FF\\ (3)\end{tabular} & \begin{tabular}[c]{@{}r@{}}\textcolor{red}{FF FF FF FF FF FF 00}\\ (3)\end{tabular} & \begin{tabular}[c]{@{}r@{}}\textcolor{red}{FF FF FF FF FF FF 00}\\ (3)\end{tabular} \\ \hline
\multirow{2}{*}{6}       & \multicolumn{1}{l|}{\multirow{2}{*}{TP.CM\_EndofMsgACK}}         & Bytes received           & 18                                                                 & 18                                                                 & 18                                                                 \\  
                         & \multicolumn{1}{l|}{}                                            & Packets received         & 3                                                                  & 3                                                                  & 3                                                                  \\ \hline
\end{tabular}
\end{adjustbox}
\end{table*}

\subsection{TP.DT Spoofing Attack}
\subsubsection{Attack Scenario and Testing}
The 14th attack scenario introduces an unknown approach proposed here in which the attacker disrupts normal communication by transmitting more data packets than the sender initially planned, leading to confusion at the receiving node.
In \Cref{fig:scenario_attack14} illustrates the TP.DT spoofing attack.
In this scenario, the legitimate ECU intends to dispatch three TP.DT messages for its communication session.
The attacker intervenes by sending additional TP.DT packets to the receiving node before the original session concludes.
The surplus packets, arriving unexpectedly at the receiver, are designed to create data overflow or processing anomalies, thereby undermining the integrity of the communication session and potentially causing incorrect data processing or system malfunctions at the receiving ECU.
This attack method exploits the communication protocol's handling of multi-packet transmissions, targeting the reliability of data transfers between ECUs.
ECU 0 is configured to send three TP.DT messages to ECU 249 in the testbed.
Following the transmission of the third TP.DT message, an additional TP.DT message is dispatched immediately.
The introduction of extra data via the immediate sending of an additional TP.DT message is anticipated to lead to a buffer overflow due to the unavailability of sufficient buffer space to accommodate the unexpected fourth packet.

\subsubsection{Results}
Following the transmission of the third TP.DT message, an injection of either the fourth TP.DT message or both the fourth and the fifth TP.DT message was performed to observe the resultant behavior. 
The session concludes with the issuance of a TP.CM\_EndofMsgACK packet signified that only three packets have been successfully transmitted, irrespective of the additional TP.DT messages transmitted.

\subsection{Analysis of Results}
Table \ref{table:proposed_attack_scenarios} summarizes the results of the above experiment.
We test fourteen attack scenarios on a sophisticated testbed configured with SAE J1939 simulators.
Attack scenarios previously successful on existing testbeds also achieved success on our testbed.
It demonstrates that our testbed provides an environment similar to those used by previous researchers.
The proposed attack scenarios exploit vulnerabilities identified within SAE J1939 protocol standard.
Specifically, we identify several scenarios where a single packet can facilitate an attack. Such attacks pose a critical threat due to their low detectability by current intrusion detection systems.

\begin{table*}[t]
\setlength{\tabcolsep}{10pt}
\centering
\caption{Comparative summary of various attack scenarios in SAE J1939.}
\label{table:proposed_attack_scenarios}
\begin{adjustbox}{max width=\linewidth}
\begin{tabular}{ll|rr|rr|rr}
\hline
\textbf{Attack scenario name}           & \textbf{\begin{tabular}[c]{@{}l@{}}Initial proposer\\ (subsequent experimenters)\end{tabular}} & \textbf{\begin{tabular}[c]{@{}r@{}}Testbed\\ experiment\end{tabular}}            & \textbf{\begin{tabular}[c]{@{}r@{}}Testbed \\ impact\end{tabular}} & \textbf{\begin{tabular}[c]{@{}r@{}}Actual\\ vehicle\\ experiment \end{tabular}}            & \textbf{\begin{tabular}[c]{@{}r@{}}Actual\\ vehicle\\ impact\end{tabular}} & \textbf{\begin{tabular}[l,b]{@{}r@{}}Our \\testbed \\ experiment\end{tabular}} & \textbf{\begin{tabular}[c]{@{}r@{}}Our\\ testbed\\ impact\end{tabular}} \\ \hline
\begin{tabular}[l,b]{@{}r@{}}Request overload attack\\ \newline \end{tabular}                 & \begin{tabular}[l,b]{@{}l@{}}Mukherjee \textit{et al.}\cite{mukherjee2016practical}\\ (Chatterjee \textit{et al.}\cite{chatterjee2023exploiting})\end{tabular}                & \begin{tabular}[l,b]{@{}r@{}}\CheckmarkBold\\ \CheckmarkBold\end{tabular} & \begin{tabular}[l,b]{@{}r@{}}\CheckmarkBold\\ \CheckmarkBold\end{tabular}                                         & \begin{tabular}[l,b]{@{}r@{}}
\\ \CheckmarkBold\end{tabular} & \begin{tabular}[l,b]{@{}r@{}}
\\ \CheckmarkBold\end{tabular}                                         & \begin{tabular}[l,b]{@{}r@{}}\CheckmarkBold\\ \newline \end{tabular}                                        & \begin{tabular}[l,b]{@{}r@{}}\CheckmarkBold\\ \newline \end{tabular}                                                                                      \\
Malicious Acknowledgment attack         & This study                                                                                           &                                                 &                                                                                         &                                                 &                                                                                         & \ding{56}                                        &                                                                                       \\
Malicious TP.CM\_RTS attack             & Mukherjee \textit{et al.}\cite{mukherjee2016practical}                                                                              &                                                 &                                                                                         &                                                 &                                                                                         & \CheckmarkBold                                        & \CheckmarkBold                                                                                       \\
\begin{tabular}[l,b]{@{}r@{}}Connection exhaustion attack\\ \newline \end{tabular}            & \begin{tabular}[l,b]{@{}l@{}}Mukherjee \textit{et al.}\cite{mukherjee2016practical}\\ (Chatterjee \textit{et al.}\cite{chatterjee2023exploiting})\end{tabular}                & \begin{tabular}[l,b]{@{}r@{}}\CheckmarkBold\\ \CheckmarkBold\end{tabular} & \begin{tabular}[l,b]{@{}r@{}}\CheckmarkBold\\ \CheckmarkBold\end{tabular}                                         & \begin{tabular}[l,b]{@{}r@{}}\\ \CheckmarkBold\end{tabular} & \begin{tabular}[l,b]{@{}r@{}}\\ \CheckmarkBold\end{tabular}                                         & \begin{tabular}[l,b]{@{}r@{}}\CheckmarkBold\\ \newline \end{tabular}                                        & \begin{tabular}[l,b]{@{}r@{}}\CheckmarkBold\\ \newline \end{tabular}                                                                                       \\
TP.CM\_BAM block attack                 & Chatterjee \textit{et al.}\cite{chatterjee2023exploiting}                                                                             & \CheckmarkBold                                               & \CheckmarkBold                                                                                       & \CheckmarkBold                                               & \ding{56}                                                                                       & \CheckmarkBold                                        & \CheckmarkBold                                                                                       \\
Memory leak using TP.CM\_CTS attack 1   & Chatterjee \textit{et al.}\cite{chatterjee2023exploiting}                                                                             & \CheckmarkBold                                               & \CheckmarkBold                                                                                       & \CheckmarkBold                                               & \ding{56}                                                                                       & \CheckmarkBold                                        & \CheckmarkBold                                                                                       \\
Memory leak using TP.CM\_CTS attack 2   & Chatterjee \textit{et al.}\cite{chatterjee2023exploiting}                                                                             & \CheckmarkBold                                               & \CheckmarkBold                                                                                       & \CheckmarkBold                                               & \ding{56}                                                                                       & \CheckmarkBold                                        & \CheckmarkBold                                                                                       \\
Memory leak using TP.CM\_CTS attack 3   & This study                                                                                           &                                                 &                                                                                         &                                                 &                                                                                         & \CheckmarkBold                                        & \ding{56}                                                                                       \\
TP.CM\_EndofMsgACK interruptions attack & This study                                                                                           &                                                 &                                                                                         &                                                 &                                                                                         & \CheckmarkBold                                        & \CheckmarkBold                                                                                       \\
TP.Conn\_Abort interruptions attack     & Murvay and Groza\cite{murvay2018security}                                                                              & \CheckmarkBold                                               & \CheckmarkBold                                                                                       &                                          &                                                                                        & \CheckmarkBold                                        & \CheckmarkBold                                                                                       \\
Out of memory using TP.CM\_BAM attack   & This study                                                                                           &                                                 &                                                                                         &                                                 &                                                                                         & \CheckmarkBold                                        & \CheckmarkBold                                                                                       \\
Out of memory using TP.CM\_RTS attack   & This study                                                                                           &                                                 &                                                                                         &                                                 &                                                                                         & \CheckmarkBold                                        & \CheckmarkBold                                                                                       \\
Overwrite memory using TP.DT attack     & This study                                                                                           &                                                 &                                                                                         &                                                 &                                                                                         & \CheckmarkBold                                        & \CheckmarkBold                                                                                       \\
TP.DT spoofing attack                   & This study                                                                                           &                                                 &                                                                                         &                                                 &                                                                                         & \CheckmarkBold                                        & \ding{56}                                          \\ \hline                               
\end{tabular}
\end{adjustbox}
\end{table*}

\section{Conclusion and Future Work} \label{sec:conclusion}

To the best of our knowledge, this is the first attempt to investigate attack scenarios while examining all parts of SAE J1939 from the perspective of weaknesses in network protocols. This study highlights vulnerabilities in SAE J1939 security and the urgent need for enhanced protection measures. By detailing specific attack scenarios, our work will be helpful in developing more robust security frameworks. 

Our future work will develop and implement practical solutions, like network-based intrusion detection systems (IDSs), which can provide real-time monitoring and detection capabilities specific to SAE J1939 protocol. Although end-to-end encryption could theoretically offer robust security, the computational demands can be challenging for the resource-constrained environments found in vehicular networks. Therefore, centralized network-based IDSs could provide a more practical and efficient approach to detecting attacks. In the future, we aim to design and implement IDSs to detect the attacks illustrated in this study, providing a feasible method to significantly enhance the cybersecurity of SAE J1939-based systems.

\newpage

\bibliographystyle{unsrt}
\bibliography{references}

\appendix
\chapter*{\raggedright Appendix}

\begin{figure}
  \centering
  \begin{subfigure}{.33\textwidth}
    \centering
    \includegraphics[width=0.9\linewidth]{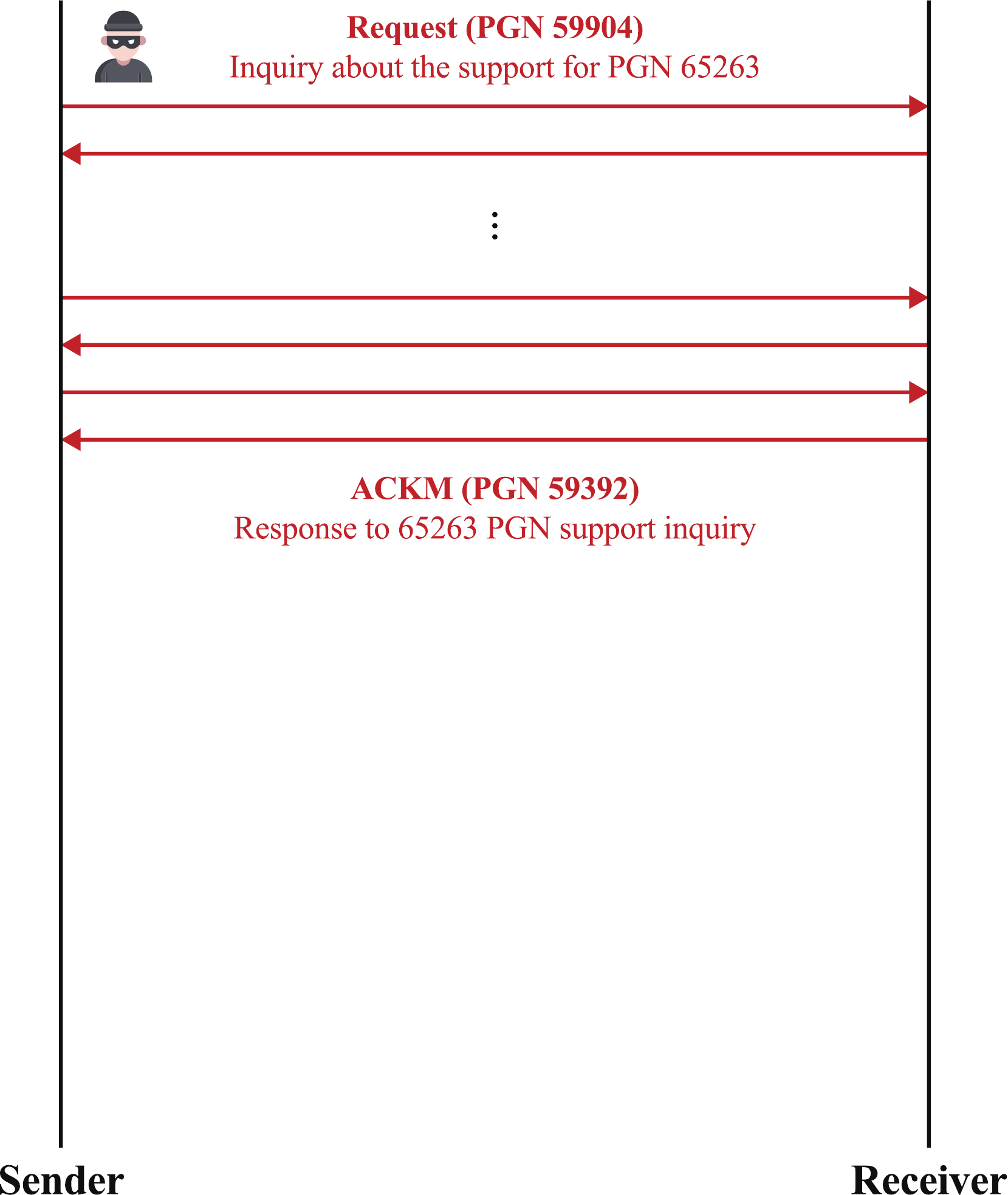}
    \caption{Request overload attack}
    \label{fig:scenario_attack1}
  \end{subfigure}%
  \begin{subfigure}{.33\textwidth}
    \centering
    \includegraphics[width=0.9\linewidth]{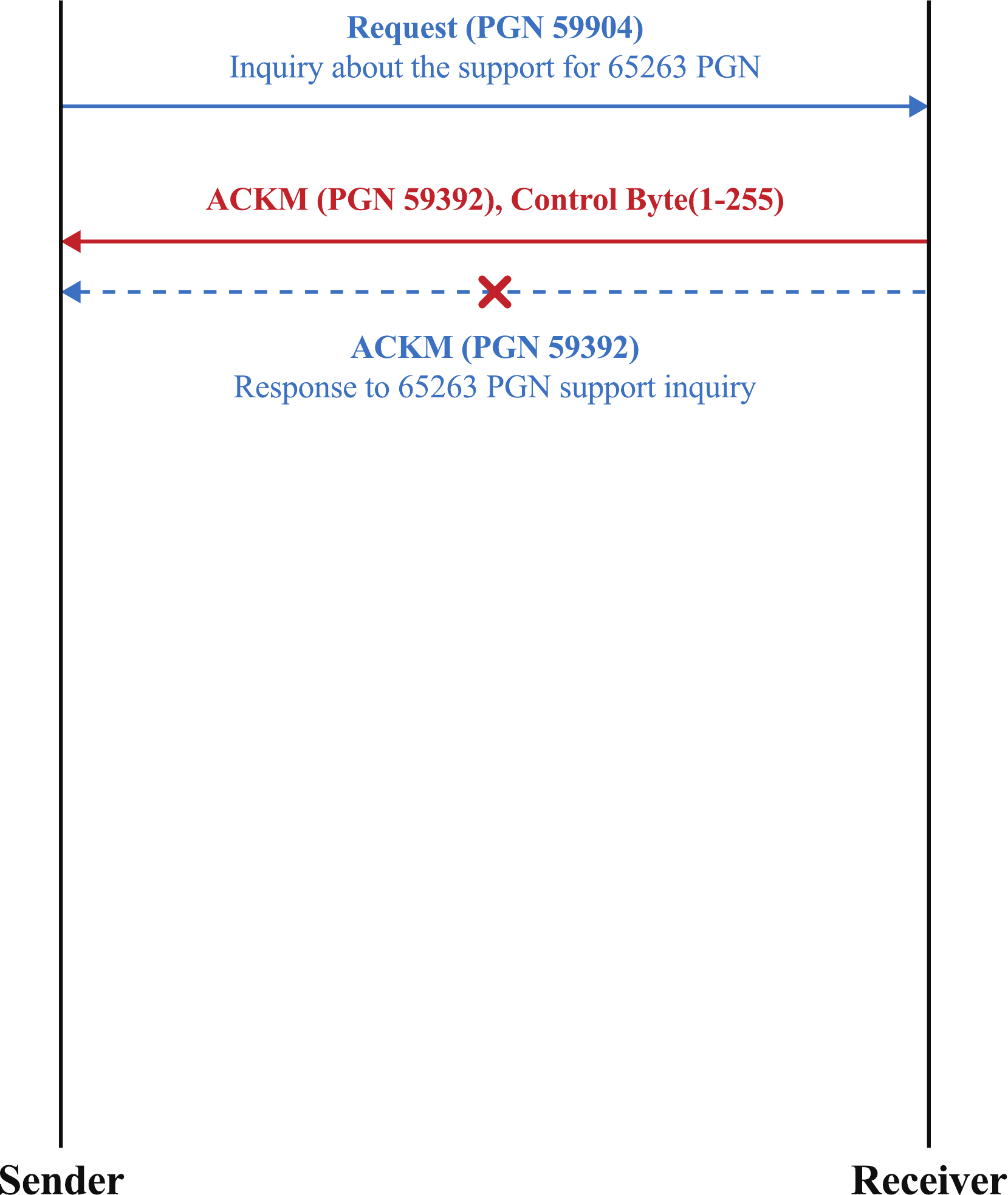}
    \caption{Malicious acknowledgment attack}
    \label{fig:scenario_attack2}
  \end{subfigure}%
  \begin{subfigure}{.33\textwidth}
    \centering
    \includegraphics[width=0.9\linewidth]{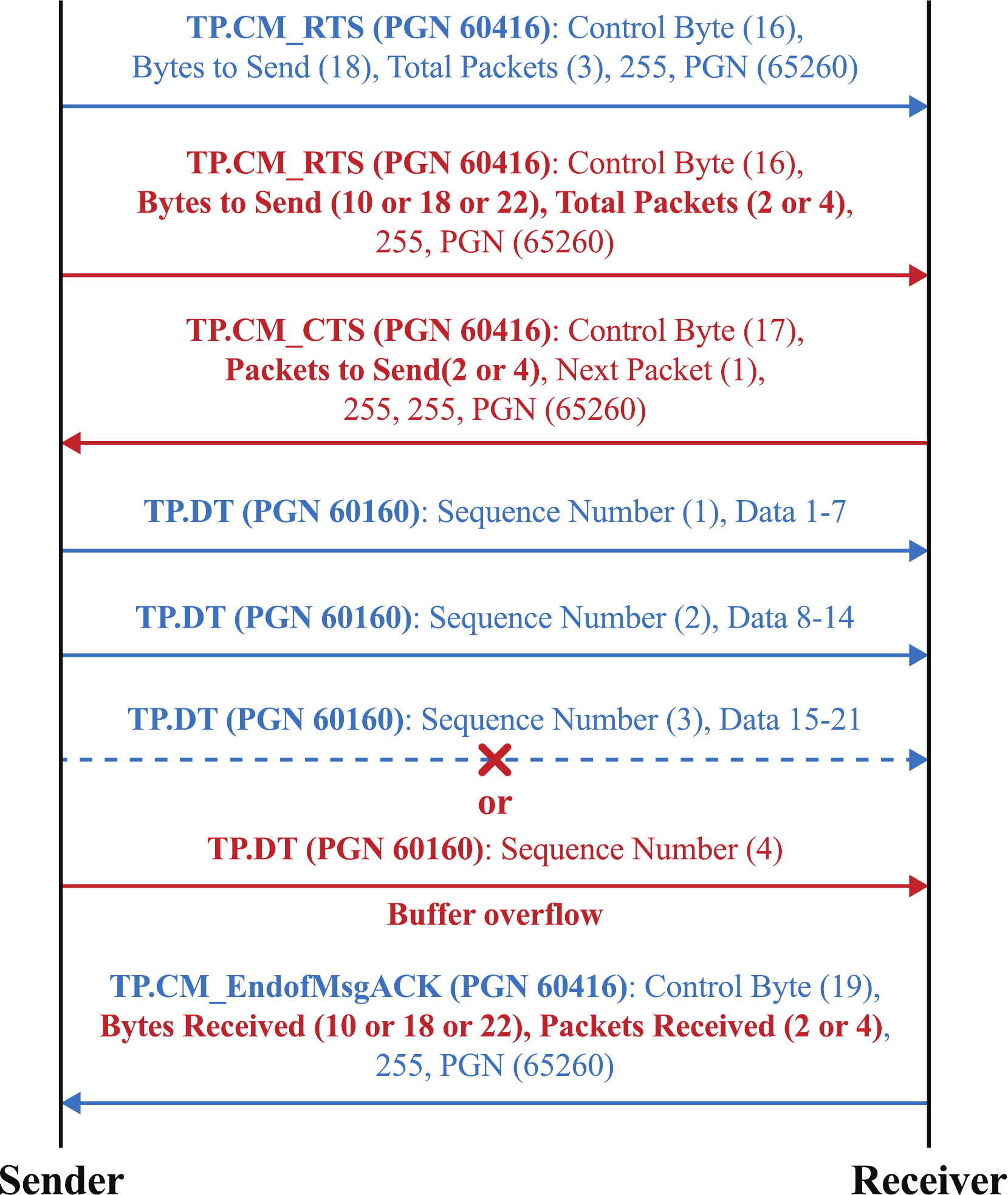}
    \caption{Malicious TP.CM\_RTS attack}
    \label{fig:scenario_attack3}
  \end{subfigure}%
  \vspace{5mm}

  \begin{subfigure}{.33\textwidth}
    \centering
    \includegraphics[width=0.9\linewidth]{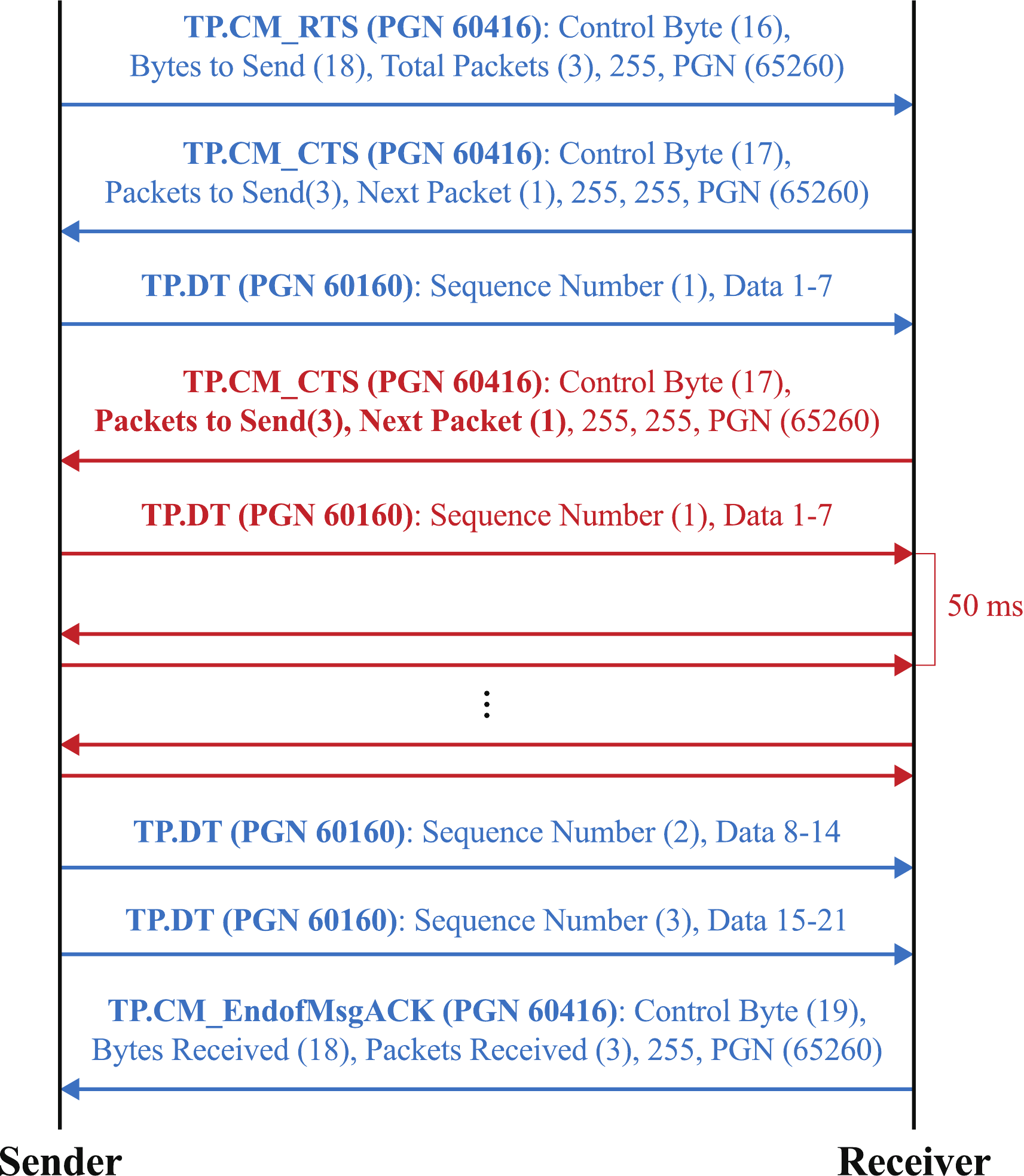}
    \caption{Connection exhaustion attack}
    \label{fig:scenario_attack4}
  \end{subfigure}
  \begin{subfigure}{.66\textwidth}
    \centering
    \includegraphics[width=0.9\linewidth]{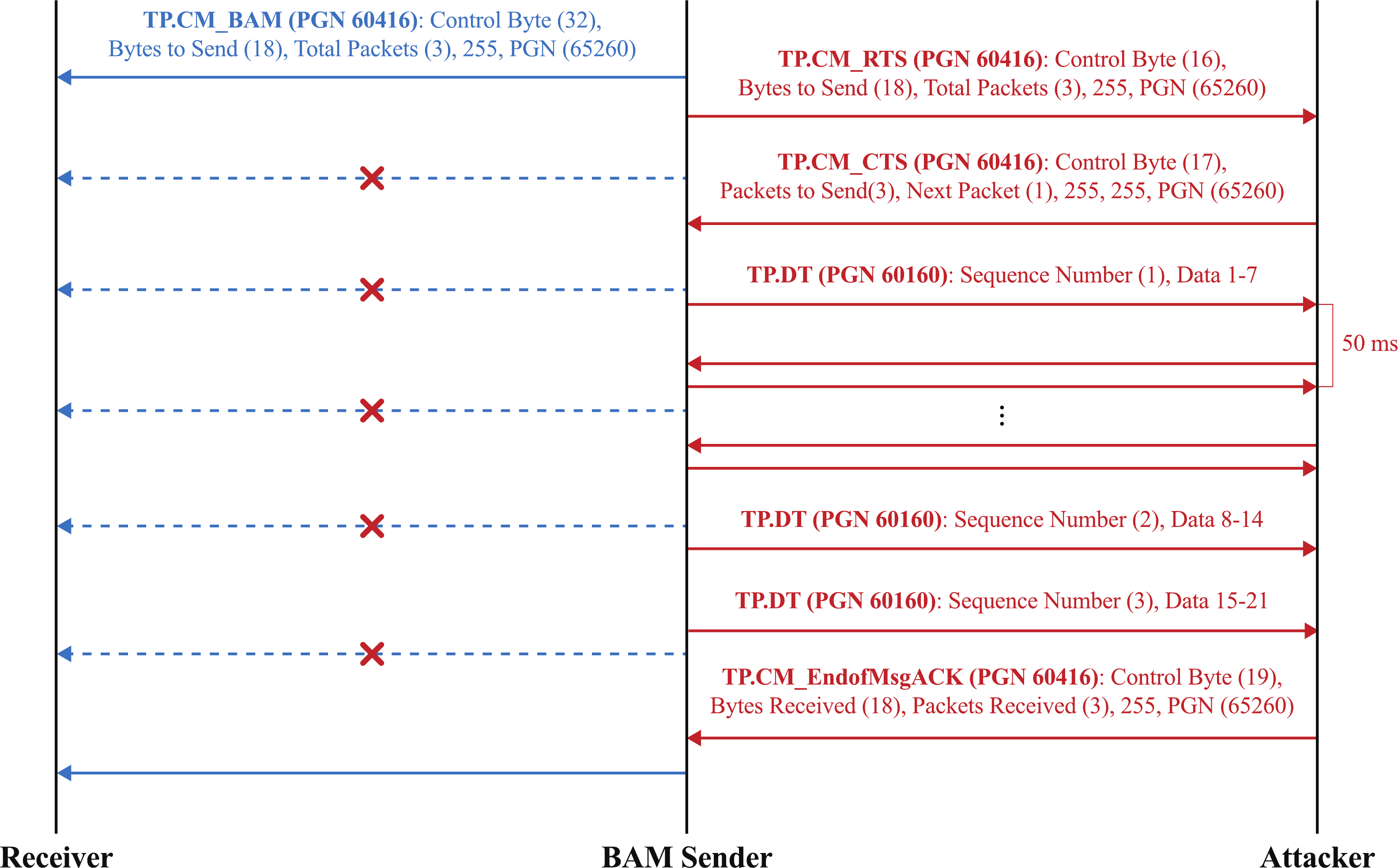}
    \caption{TP.CM\_BAM block attack}
    \label{fig:scenario_attack5}
  \end{subfigure}%
  \vspace{5mm}
\end{figure}
\clearpage

\begin{figure}[!t]
  \ContinuedFloat
  \centering
  \begin{subfigure}{.33\textwidth}
    \centering
    \captionsetup{justification=centering}
    \includegraphics[width=0.9\linewidth]{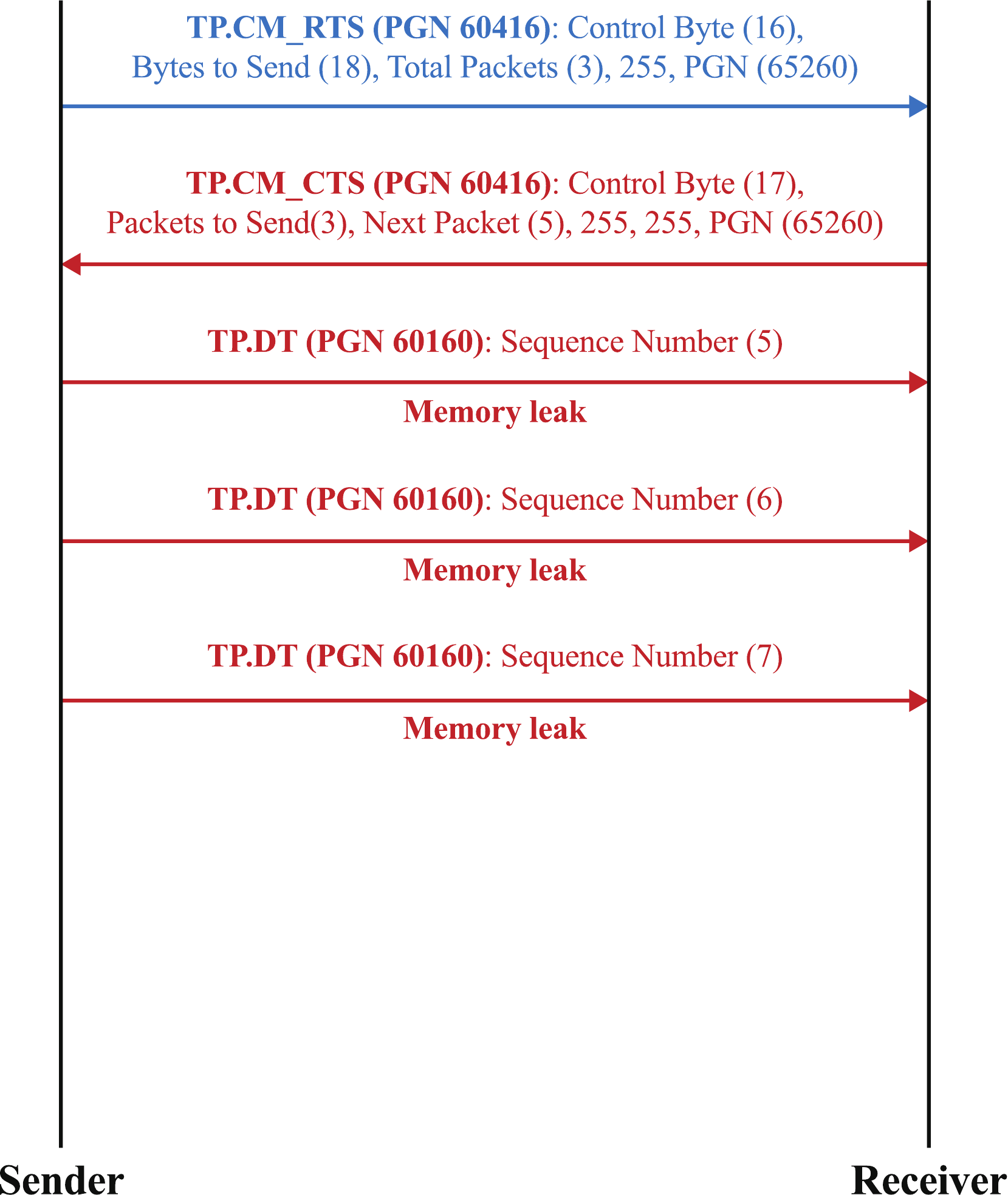}
    \caption{Memory leak using\\TP.CM\_CTS attack 1}
    \label{fig:scenario_attack6}
  \end{subfigure}%
  \begin{subfigure}{.33\textwidth}
    \centering
    \captionsetup{justification=centering}
    \includegraphics[width=0.9\linewidth]{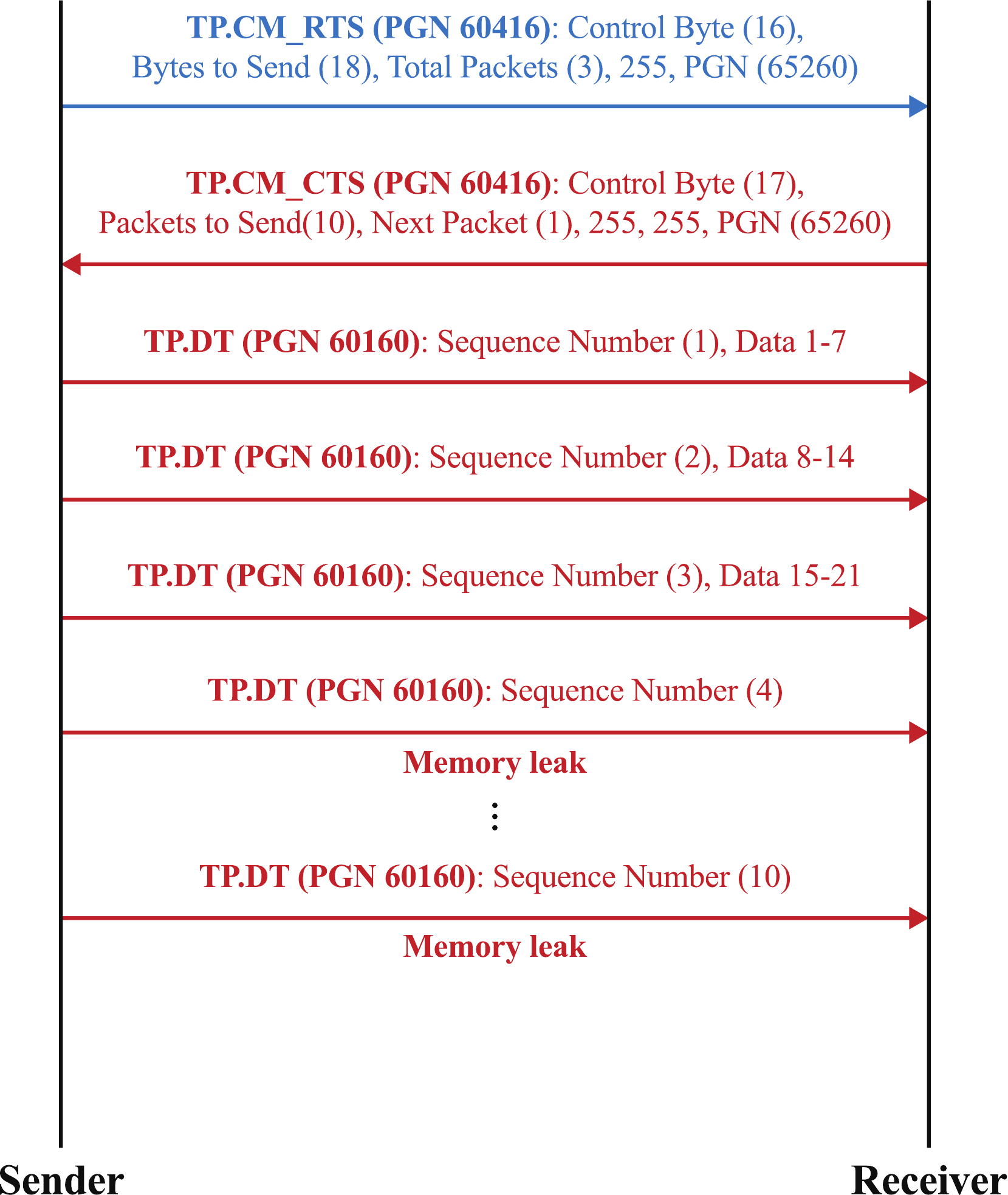}
    \caption{Memory leak using\\TP.CM\_CTS attack 2}
    \label{fig:scenario_attack7}
  \end{subfigure}%
  \begin{subfigure}{.33\textwidth}
    \centering
    \captionsetup{justification=centering}
    \includegraphics[width=0.9\linewidth]{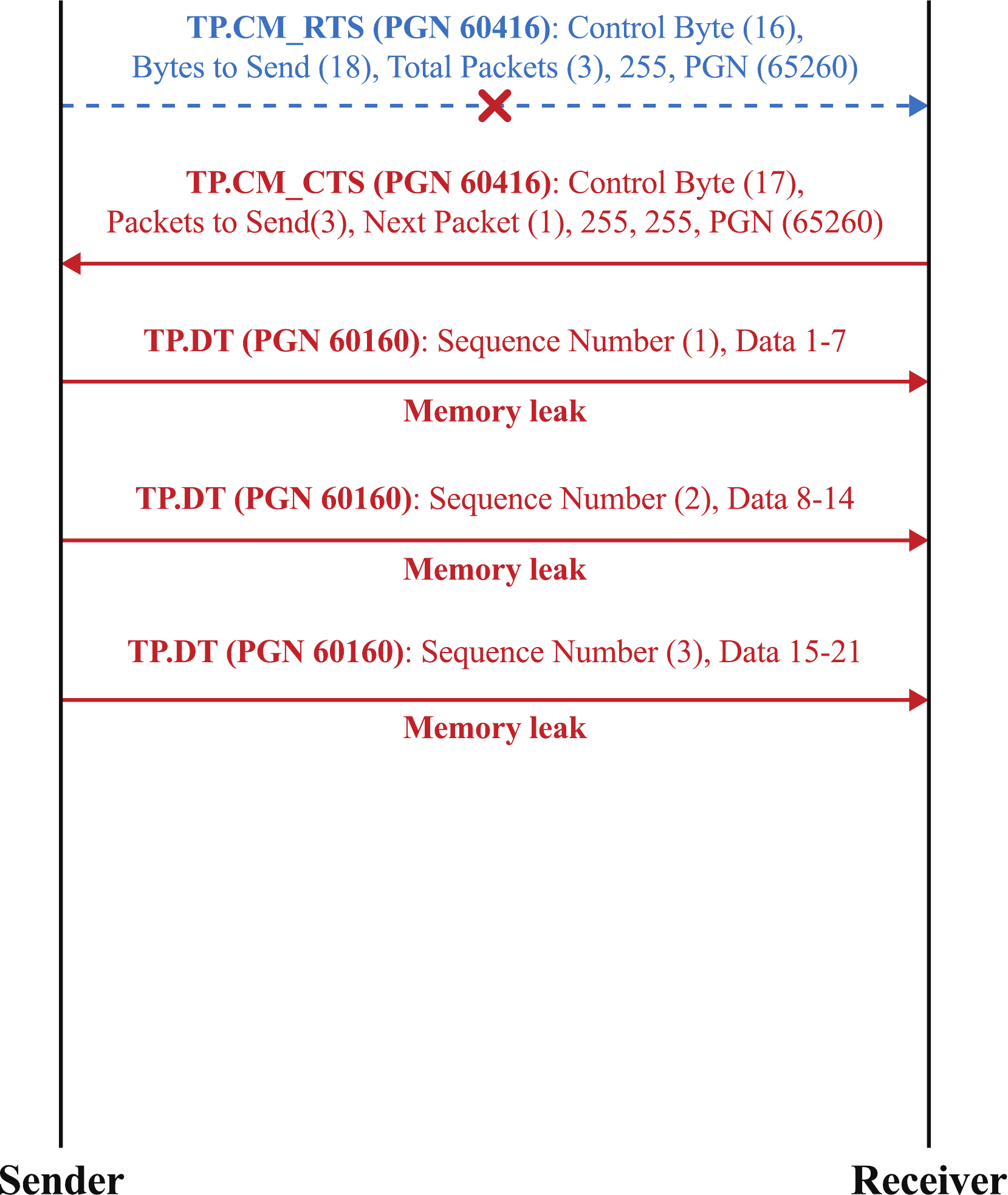}
    \caption{Memory leak using\\TP.CM\_CTS attack 3}
    \label{fig:scenario_attack8}
  \end{subfigure}
  \vspace{5mm}

  \begin{subfigure}{.33\textwidth}
    \centering
    \captionsetup{justification=centering}
    \includegraphics[width=0.9\linewidth]{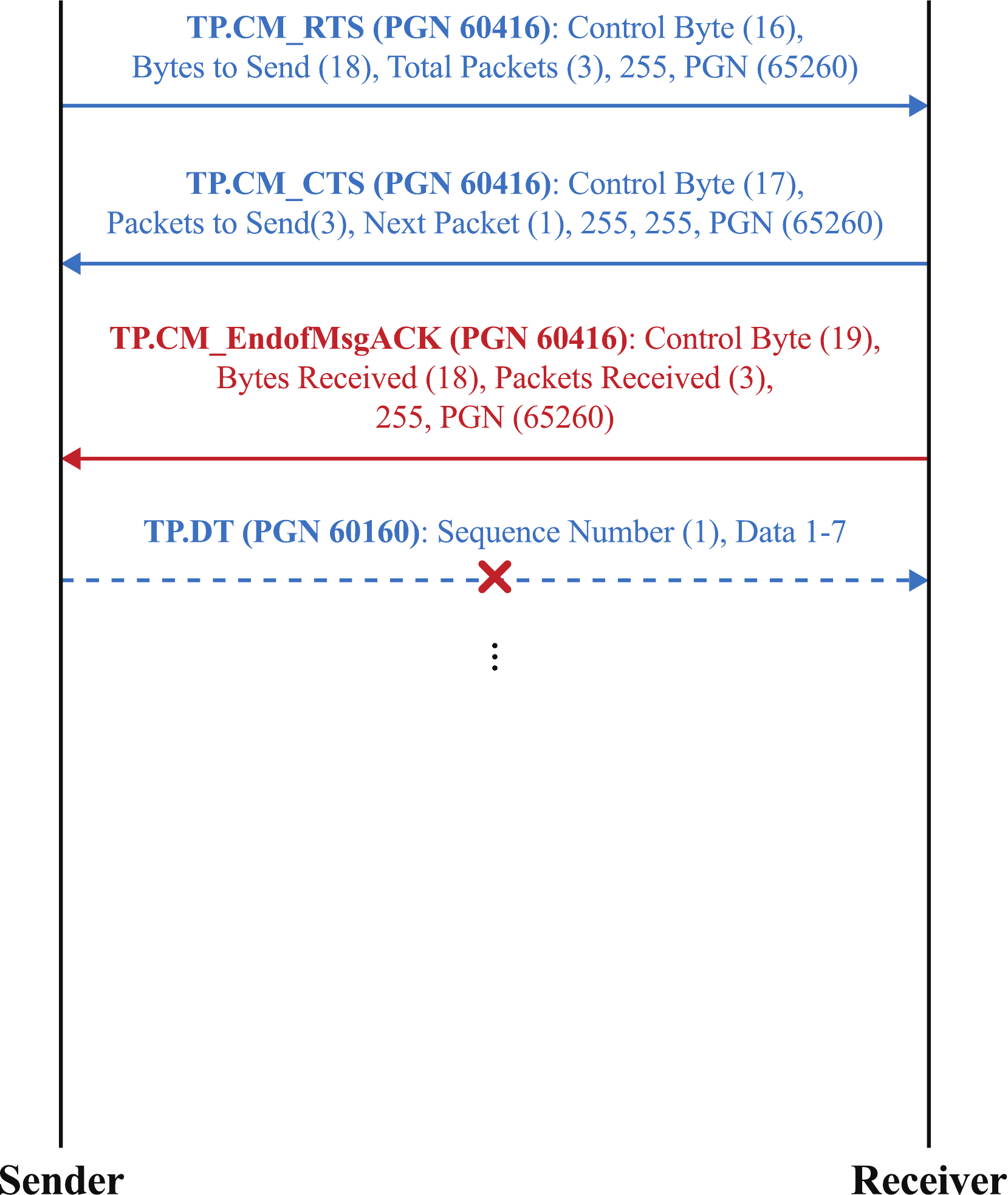}
    \caption{TP.CM\_EndofMsgACK\\interruptions attack}
    \label{fig:scenario_attack9}
  \end{subfigure}%
  \begin{subfigure}{.33\textwidth}
    \centering
    \captionsetup{justification=centering}
    \includegraphics[width=0.9\linewidth]{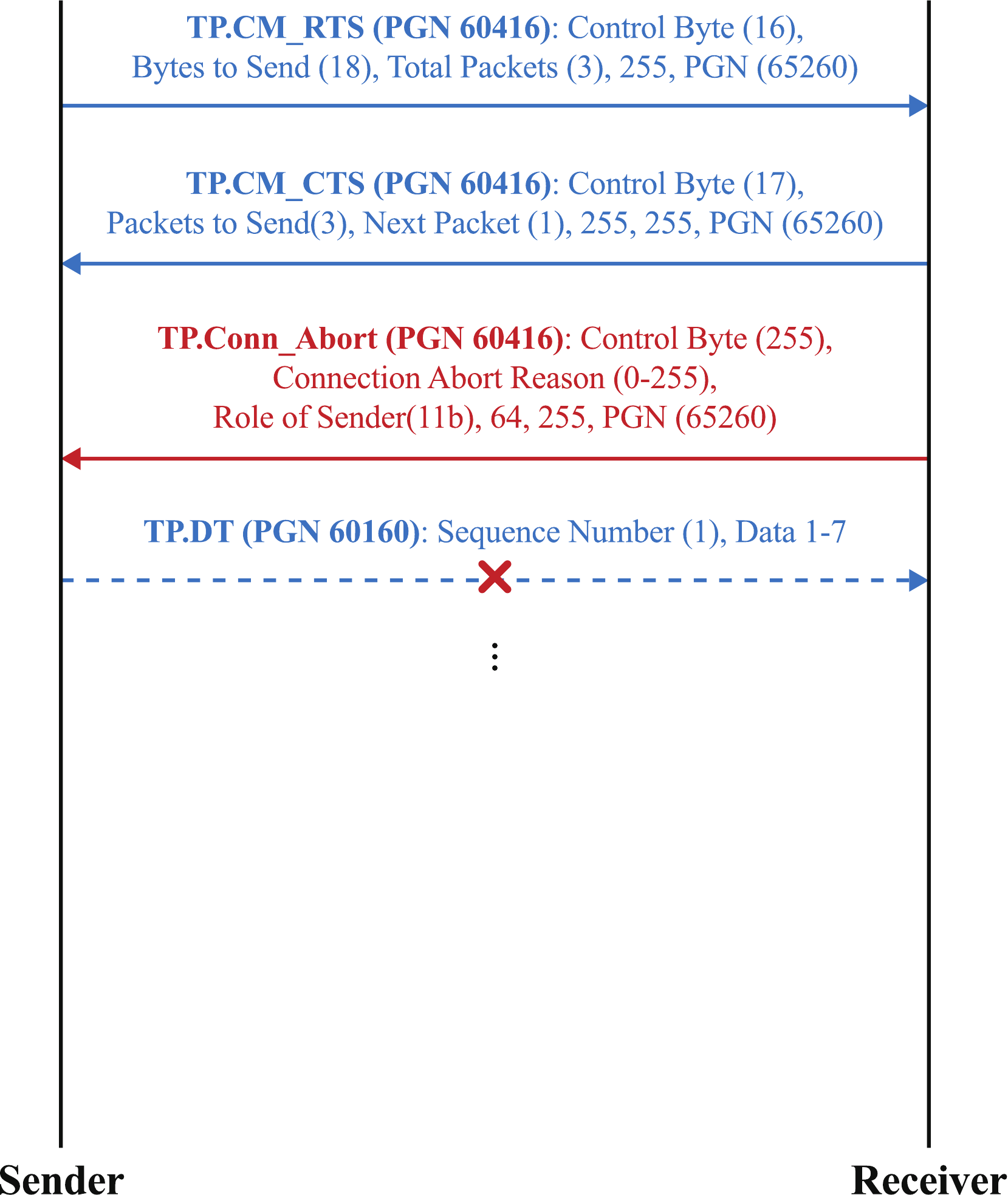}
    \caption{TP.Conn\_Abort\\interruptions attack}
    \label{fig:scenario_attack10}
  \end{subfigure}%
  \begin{subfigure}{.33\textwidth}
    \centering
    \captionsetup{justification=centering}
    \includegraphics[width=0.9\linewidth]{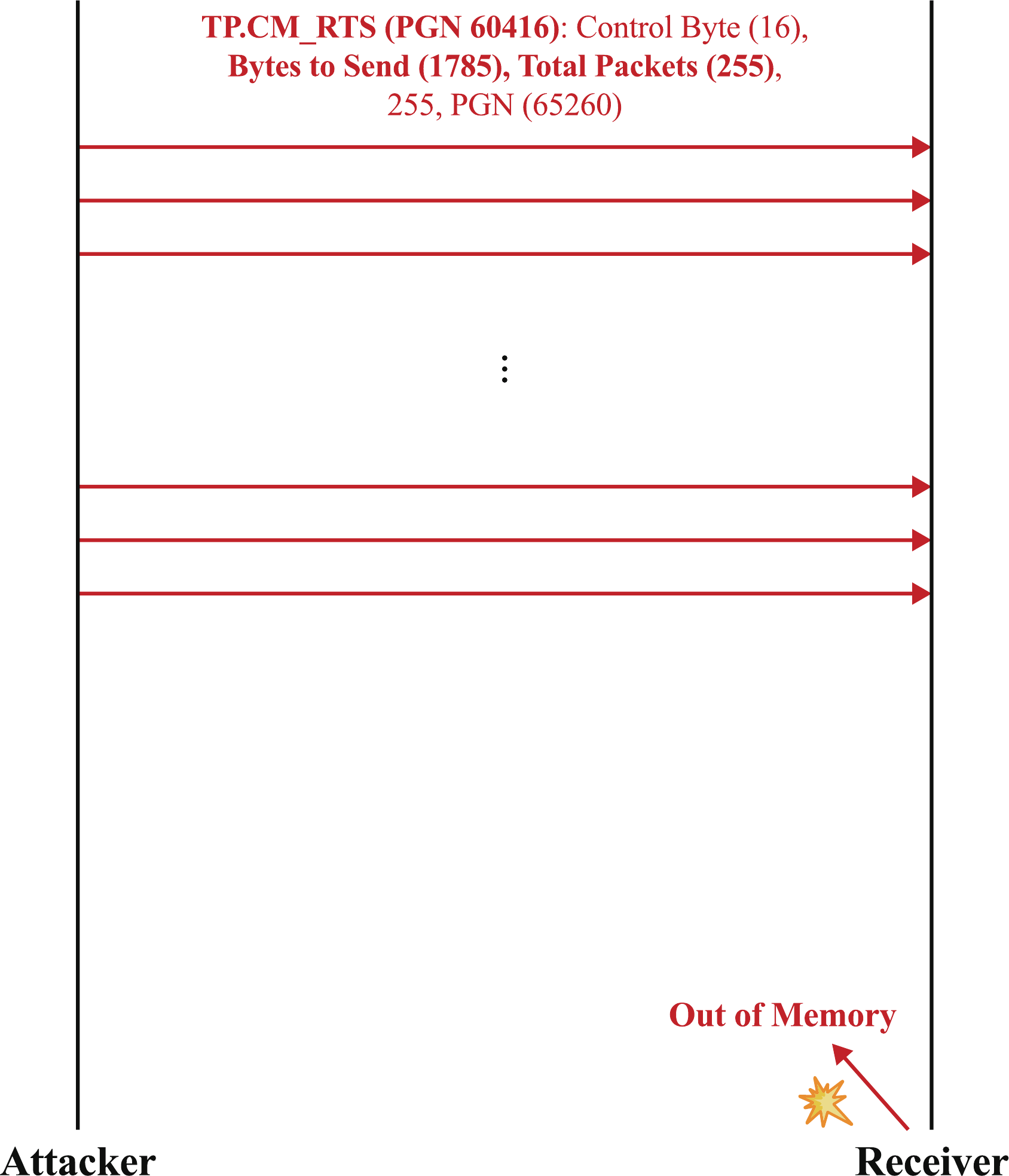}
    \caption{Out of memory using\\TP.CM\_RTS attack}
    \label{fig:scenario_attack12}
  \end{subfigure}
  \vspace{5mm}

  \caption{Explanations for 11 out of the 14 discovered attack scenarios not illustrated in \Cref{fig:attack_scenarios}}
  \label{fig:appendix_attack_scenarios}
\end{figure}
\end{document}